\documentclass[aps,prc,reprint,showpacs,superscriptaddress,twocolumn,notitlepage]{revtex4-1}

\usepackage{graphicx,color} 
\usepackage{bm}       
\usepackage{bbm}
\usepackage{amsmath} 
\usepackage[dvipsnames]{xcolor}
\usepackage{ulem}
\usepackage{multirow}
\usepackage{float} 
\newcommand {\nc} {\newcommand}

\nc{\half}{\frac{1}{2}}
\nc{\numberthis}{\addtocounter{equation}{1}\tag{\theequation}}

\nc{\scrK}{\mathcal{K}} 
\nc{\pow}[2][2]{^{\hspace{#2pt}#1}} 
\nc{\prel}{p_\text{rel}}
\nc{\qsum}{Q_\text{sum}}
\nc{\elab}{E_\text{lab}}
\nc{\meff}{m_\text{eff}}
\nc{\yref}{y_\text{ref}}
\nc{\np}{\textit{np}}
\nc{\pp}{\textit{pp}}
\nc{\npref}{\text{Im}(A_{np}^{(5)})}
\nc{\ppref}{\text{Im}(A_{np}^{(5)})}

\nc{\Renp}[1]{$\Re$e$\,{#1}_{np}$}
\nc{\Imnp}[1]{$\Im$m$\,{#1}_{np}$}
\nc{\Repp}[1]{$\Re$e$\,{#1}_{pp}$}
\nc{\Impp}[1]{$\Im$m$\,{#1}_{pp}$}

\nc{\Ree}[1]{$\Re$e$\,{#1}$}
\nc{\Imm}[1]{$\Im$m$\,{#1}$}

\nc{\halfwidth}{0.48\textwidth}
\nc{\MDwidth}{0.25\linewidth}
\newlength{\prcColumnWidth} 
\newlength{\cnWidth} 
\newlength{\mdWidth} 

\nc {\IR} [1]{\textcolor{red}{#1}}
\nc {\IB} [1]{\textcolor{blue}{#1}}         
\nc {\IP} [1]{\textcolor{magenta}{#1}}
\nc {\IM} [1]{\textcolor{Bittersweet}{#1}}  
\nc {\IE} [1]{\textcolor{Plum}{#1}}         
\nc {\IG} [1]{\textcolor{OliveGreen}{#1}}

\begin{document}

\title{Order-by-order uncertainties of nucleon-nucleon Wolfenstein amplitudes in chiral effective field theory}

\author{B.~McClung}
\affiliation{Institute of Nuclear and Particle Physics, and Department of Physics and Astronomy, Ohio University, Athens, OH 45701, USA}

\author{Ch.~Elster}
\affiliation{Institute of Nuclear and Particle Physics, and Department of Physics and Astronomy, Ohio University, Athens, OH 45701, USA}

\author{D.R.~Phillips}
\affiliation{Institute of Nuclear and Particle Physics, and Department of Physics and Astronomy, Ohio University, Athens, OH 45701, USA}

\date{\today}

\begin{abstract}

Quantum mechanical invariance principles dictate the most general operator structure that can be present in the nucleon-nucleon (NN) interaction. Five independent operators appear in the on-shell NN amplitude together with five corresponding coefficient functions. The usual choice for these coefficient functions is known as the NN Wolfenstein amplitudes. 
We analyze the order-by-order convergence of each of the five NN Wolfenstein amplitudes predicted by a semi-local coordinate space potential implementation of chiral effective field theory ($\chi$EFT). We do this at laboratory kinetic energies between 25 and 200~MeV for both neutron-proton and proton-proton scattering. Our analysis uses the Gaussian-Process methods developed by the BUQEYE collaboration to describe the contributions of each $\chi$EFT order, and so yields truncation uncertainties for each Wolfenstein amplitude that are correlated across scattering angles. We combine information on the size of different orders in the EFT to infer the $\chi$EFT breakdown scale for each amplitude, finding, on average, $\Lambda_b$ between 750 and 800~MeV. 
With this choice of $\Lambda_b$, the EFT truncation uncertainties cover both higher-order results and empirical Wolfenstein amplitudes well for all orders other than the leading order.  


\end{abstract}


\maketitle

\section{Introduction and Motivation}
\label{sec:intro}

Nucleon-nucleon (NN) interactions derived in the chiral effective field theory ($\chi$EFT) expansion are widely applied in 
low-energy nuclear physics. (For  reviews see Refs.~\cite{Epelbaum:2008ga,Hammer:2019poc,machleidt2024recent}.) They are crucial inputs to {\it ab initio} calculations of finite
nuclei, nuclear reactions, and nuclear matter. Such treatments of the NN force are systematic because of the effective field theory (EFT) power
counting, which organizes the infinite set of operators that could appear in the NN potential into a sequence of terms of decreasing
importance. Such a treatment implies that NN scattering observables will have a well-defined uncertainty corresponding to the size of the omitted terms beyond the order computed in the $\chi$EFT expansion. However, while in principle this statement provides a way to estimate the theory uncertainty of a $\chi$EFT calculation, quantifying those uncertainties requires a statistical model for the higher-order terms. A careful check of the uncertainty quantification of NN amplitudes derived in $\chi$EFT is also important because 
the expansion of the NN potential does not translate into an expansion of the NN amplitude. Hence, testing whether the $\chi$EFT expansion has a regular convergence pattern is an important {\it a posteriori} check on attempts to implement a $\chi$EFT expansion for observables via a NN potential.

The study of truncation uncertainties of two-nucleon scattering observables obtained with $\chi$EFT NN interactions has already been  successfully carried out 
(see e.g.~\cite{Furnstahl:2015rha,Melendez:2017phj,Melendez:2019izc}),
extended to the nucleon-deuteron (Nd)
system~\cite{Epelbaum:2019zqc}, as well as to structure observables for light
nuclei~\cite{Binder:2018pgl,Maris:2020qne,LENPIC:2022cyu}, to nucleon-nucleus scattering~\cite{Baker:2021iqy}, and to neutron and nuclear matter~\cite{Drischler:2020hwi,Drischler:2020yad}.
In a recent work Millican {\it et al.}~\cite{Millican:2024yuz} assessed correlated truncation errors
of a given $\chi$EFT NN potential in terms of NN observables, i.e., cross sections and spin observables, as functions of the laboratory kinetic energy. 
Since the observables for the NN system are plentiful, it is not always easy to relate them to specific properties of the underlying NN potentials.

In this work, we follow the same statistical methodology as Millican {\it et al.} and several of the other works listed in the previous paragraph: the ``BUQEYE" (Bayesian Uncertainty Quantification: Errors for Your EFT) approach to truncation errors. But here, we emphasize the invariant operator structure of the NN
potential~\cite{Okubo:1958qej} and the resulting scattering amplitudes.  Representing NN potentials in terms of their invariant operator
structure has already been successfully applied in calculations of NN scattering in which conventional partial wave decompositions were not
used~\cite{Veerasamy:2012sp,Golak:2010wz,Fachruddin:2000wv}. Since we are considering NN scattering, 
we will focus on the convergence pattern of the NN scattering amplitude as expressed through its  Wolfenstein 
representation~\cite{wolfenstein-ashkin,Wolfenstein:1956xg}. This means we consider five (complex) functions, which are the coefficients in an expression that defines the 
NN t-matrix in terms of the most general operator structure consistent with rotational invariance,
parity, and time reversal. 
Quantum mechanics mandates that observables are bilinear in the Wolfenstein amplitudes, so
the pattern of convergence of the chiral EFT expansion is more straightforward when analyzed in the amplitudes 
 rather than in terms of observables. 

The BUQEYE approach to EFT truncation errors involves expressing the quantities of interest in terms of dimensionless coefficients associated with each order in the EFT expansion. The EFT will then have regular convergence and predictable truncation uncertainties if all coefficients are of similar size. In addition, the formalism of Melendez et al.~\cite{Melendez:2019izc} allows the analysis of functions--like the Wolfenstein amplitudes--predicted by the EFT, since it assumes that each coefficient function is a different draw from the same underlying Gaussian process (GP)~\cite{rasmussen2006gaussian,BastosDiagnosticsGaussianProcess2009}. 

Here, we carry out such an analysis of the neutron-proton and proton-proton Wolfenstein amplitudes predicted in a semi-local co-ordinate space (SCS) implementation of the chiral EFT NN interaction of Epelbaum, Krebs, and Mei\ss ner~\cite{Epelbaum:2014efa,Epelbaum:2014sza}. This potential exists at five different EFT orders ranging from 
leading order LO [$O(Q^0)$], to next-to-next-to-next-to-next-to-leading order N$^4$LO [$O(Q^5)$], and so we have a sizable set of coefficients to train our GP on for each of the five Wolfenstein amplitudes we consider.

The BUQEYE formalism also allows the extraction of a posterior probability density for the breakdown scale of the EFT, i.e., the momentum scale at which all terms in the EFT have the same size.
Thus, we can extract breakdown scales associated with each Wolfenstein amplitude. However, we will also see that the pattern of EFT convergence is irregular for some of the amplitudes. 

The paper is structured as follows. In Sec.~\ref{sec:formal}, we first briefly introduce Wolfenstein amplitudes, then review basic concepts of GPs to analyze the convergence of EFT predictions and give specific choices implemented in this work. In Sec.~\ref{sec:pn}, we analyze the neutron-proton Wolfenstein amplitudes as predicted by one SCS $\chi$EFT interaction, and in Sec.~\ref{sec:pp}, we analyze the corresponding proton-proton amplitudes. We conclude in Sec.~\ref{sec:conclusions}. In an appendix, we show additional results for amplitudes that may interest some readers but might distract from the main points of our study if presented in the body of the paper.


\section{Formal Considerations}
\label{sec:formal}

\subsection{Wolfenstein amplitudes}

\label{subsec:Wolfensteins}

Efforts to constrain the NN scattering matrix $\overline{M}$ from invariance principles have a long
tradition~\cite{Hoshizaki:1968wi,Bystricky:1976jr}.  In addition to Galilean invariance, time reversal, and parity conservation are considered, and a minimal operator basis for the momentum-space amplitudes is derived.
For the scattering of two spin-1/2 particles, these 
considerations lead to five invariant amplitudes, known as Wolfenstein amplitudes, from which 25 possible experimental situations can be constructed. There are also separate amplitudes for neutron-proton ($np$) and proton-proton ($pp$) scattering---or equivalently, in the isospin limit, for the isospin-zero and isospin-one channels of NN scattering. 

We use the following definition for the NN scattering amplitude, expressed in the basis and notation of Ref.~\cite{gloecklebook}:
\begin{eqnarray}
\overline{M}(q;E) &=& A(q;E) \; \mathbbm{1} \nonumber \\
 &+& i C(q;E) (\bm{\sigma}_1 + \bm{\sigma}_2) \cdot \bm{\hat{n}} \nonumber \\
 &+& M(q;E) (\bm{\sigma}_1 \cdot \bm{\hat{n}}) (\bm{\sigma}_2 \cdot \bm{\hat{n}}) \\
 &+& [G(q;E) - H(q;E)] (\bm{\sigma}_1 \cdot \bm{\hat{q}}) (\bm{\sigma}_2 \cdot \bm{\hat{q}}) \nonumber \\
&+& [G(q;E) + H(q;E)] (\bm{\sigma}_1 \cdot \bm{\hat{\mathcal{K}}}) (\bm{\sigma}_2 \cdot 
  \bm{\hat{\mathcal{K}}}). \nonumber
\label{eq:1}
\end{eqnarray}
The momenta appearing in this operator basis are 
\begin{eqnarray}
\bm{q}& =& \bm{p'}-\bm{p} \nonumber \\
\bm{\mathcal{K}} &=& \frac{1}{2}(\bm{p'}+\bm{p}),
\end{eqnarray}
where $\bm{p}$ and $\bm{p}'$ are the relative momenta of the nucleons before and after the scattering, and thus $\bm{q}$ is the momentum transfer and $\bm{\mathcal{K}}$ is the momentum average. Unit vectors are then written as $\bm{\hat{q}}$ and $\bm{\hat\mathcal{K}}$, and the unit vector $\bm{\hat{n}}$ is perpendicular to both $\bm{\hat{q}}$ and $\bm{\hat\mathcal{K}}$: $\bm{n} \equiv \bm{p}\times\bm{p'}$. The spin operators of the two nucleons are given by $\bm{\sigma}_1$ and $\bm{\sigma}_2$.

\begin{figure}[hbt]
\includegraphics[width=0.47\textwidth] {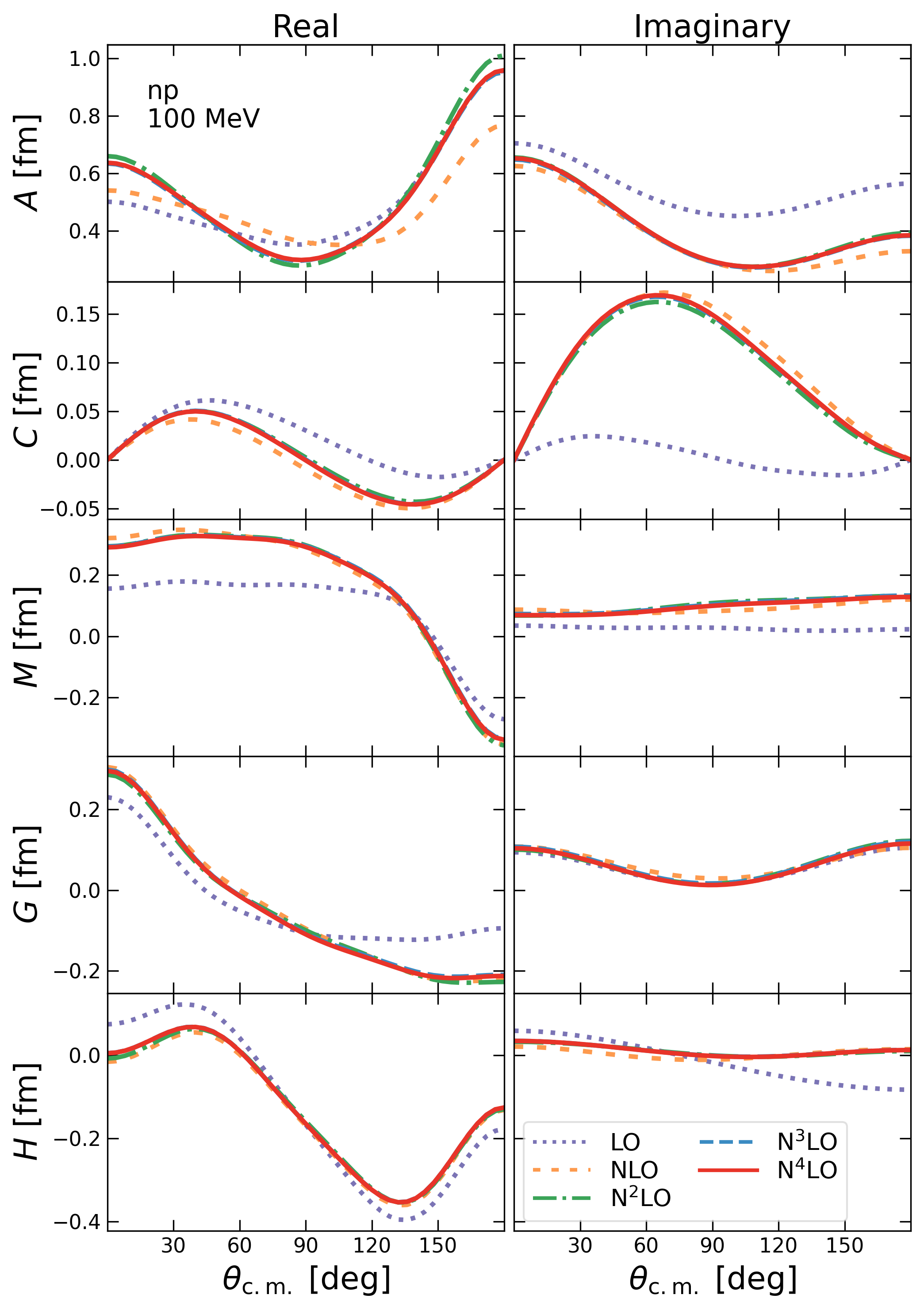}
\caption{Wolfenstein amplitudes $A$, $C$, $M$, $G$, and $H$ as  function of the c.m. scattering angle for $np$ scattering at 100~MeV laboratory kinetic energy. All calculations are based on the
SCS~\protect\cite{Epelbaum:2014efa,Epelbaum:2014sza} implementation of the $\chi$EFT NN interaction with a cutoff of $R=0.9$~fm. Results at different $\chi$EFT orders are indicated by the different line types defined in the legend.} 
\label{fig1}
\end{figure}

The Wolfenstein amplitudes $A(q;E)$, $C(q;E)$, $M(q;E)$, $G(q;E)$, and $H(q;E)$ are complex-valued functions of the magnitude of the momentum transfer (or equivalently of the scattering angle) at a given energy $E$, which can either be the laboratory kinetic energy or the center-of-mass (c.m.) energy. They multiply the invariant operators of the NN system, which are of scalar ($A$), vector ($C)$, and tensor characters ($M$, $G$, and $H$). The amplitude $C$ sums all spin-orbit contributions and includes a factor $i$, making $\Im$m$\,C$ the dominant spin-orbit contribution, while $\Re$e$\,C$ is small compared to the other amplitudes. Meanwhile, effects directly attributable to the tensor portion of the NN force appear in 
$M$, $G$, and $H$. The real parts of the tensor amplitudes $M$, $G$, and $H$  are considerably larger than their imaginary parts; the imaginary component of $H$ is especially small. $\Re\mbox{e}\,C$ and $\Im\mbox{m}\,H$ are small for all the energies we consider, so we do not examine them in our study of the order-by-order $\chi$EFT uncertainties.

Since a representation of an NN interaction in terms of Wolfenstein amplitudes may not be as familiar to readers as a representation of the interaction's amplitude in terms of partial-wave phase shifts, we show in Fig.~\ref{fig1}
the  Wolfenstein amplitudes for the different orders of the SCS $\chi$EFT interaction~\cite{Epelbaum:2014efa,Epelbaum:2014sza} for a cutoff of $R = 0.9$~fm and $np$ scattering at 100~MeV laboratory kinetic energy. Figure~\ref{fig1} shows the convergence pattern of  Wolfenstein amplitudes computed with the SCS $\chi$EFT interaction, beginning with LO, through all subsequent orders, until the highest SCS order of N$^4$LO is reached.

\subsection{Analyzing the convergence of EFT predictions using Gaussian processes}
\label{subsec:GPs}

The inputs to our analysis of the convergence of $\chi$EFT are order-by-order predictions for the five Wolfenstein amplitudes. We analyze the convergence of these amplitudes at several discrete laboratory kinetic energies.

For a particular energy, we re-express the predictions of the Wolfenstein amplitudes at each order in terms of a coefficient function, $c_n$, that defines the size and momentum-transfer dependence of that contribution:
\begin{equation}
    c_n(q;E) \equiv \frac{y_n(q;E)-y_{n-1}(q;E)}{y_{\rm ref}(q;E) Q^n},
\end{equation}
with $y_{\rm ref}(q;E)$ being a reference amplitude that captures the broad structure of all Wolfenstein amplitudes and $Q$ the expansion parameter of the EFT (cf. Eq.~(\ref{eq:Qchoice}) below). Note that $Q$ itself is momentum-transfer- and energy-dependent.

The full Wolfenstein amplitude at order $k$ is then re-constructed as
\begin{equation}
    y_k(q;E)=y_{\rm ref}(q;E)\sum_{n=0}^k c_n(q;E) Q^n.
\end{equation}
Therefore, if the $c_n$'s can be characterized in a statistical fashion we can estimate the uncertainty associated with the $k+1$st and higher terms in the $\chi$EFT series for the Wolfenstein amplitude $y$.

We follow Ref.~\cite{Melendez:2019izc}
and hypothesize that the curves $c_n(q;E)$ are all independent but identically distributed draws from a Gaussian process,
\begin{equation}
    c_n(q;E) \stackrel{\rm{i.i.d}}{\sim} \mathcal{GP}\left[0,\,\bar{c}(E) ^2r(q,q';\ell_q(E))\right].
    \label{eq:BUQEYE}
\end{equation}
The Gaussian process defines the curves $c_n$ as draws from multi-variate Gaussian probability distributions. The hypothesis (\ref{eq:BUQEYE}) means that the curves at all orders should have the same conditional probability distribution, which we assume has a mean of zero, a squared exponential correlation matrix $r(q,q';\ell_q)$ (corresponding to a correlation length $\ell_q$), and a variance $\bar{c}^2$. The resulting probability distribution for a coefficient vector $\vec{c}_n$, obtained by computing $c_n$ at some set of momenta, is given by
\begin{equation}
    {\rm pr}(\vec{c}_n) \propto \exp\left(-\frac{1}{2} \vec{c}_n^{\, T} K^{-1} \vec{c}_n\right),
\end{equation}
where the entries of the covariance matrix $K$ are defined according to the momentum transfers $q$ at which the entries in the vector $\vec{c}_n$ are evaluated:
\begin{equation}
    K(q',q;E)=\bar{c}^2(E) \exp\left[-\frac{(q'-q)^2}{2 \ell_q(E)}\right].
\end{equation}
The parameters $\bar{c}^2$ and $\ell_q$ are estimated based on the training data supplied, i.e., the discrete sets of points we choose from each $c_n$ to represent the curve.

These training data also define the conditional probability distributions that describe the curves $c_n$ at any point $q$ that is not part of the training set. Thus, while all coefficients $c_n$ have common statistical properties---a common, typical size and a common, typical scale of variation---the curves may  (and generally do) all have different values since the conditional probability distribution of points outside the training data is affected by the points chosen for that set. The ``updating equations'' that give the mean and variance associated with that conditional probability distribution of every other point on the curve can be found in Ref.~\cite{rasmussen2006gaussian}.

The fact that these equations should describe all remaining points on the coefficient curves allows us to check whether the GP hypothesis (\ref{eq:BUQEYE}) is statistically consistent. The Mahalanobis distance provides a straightforward overall check of this consistency
\begin{equation}
{D^{(n)}_{MD}}^2 \equiv \sum_{i,j=1}^{N_{\rm test}} c_n(q_i) K^{-1}(q_i,q_j) c_n(q_j),
\end{equation}
where the sum runs over a chosen set of testing points $q_i$, $i=1, \ldots, N_{\rm test}$. This is the generalization of the $\chi^2$ statistic to correlated data, and it should be distributed as a $\chi^2$ with $N_{\rm test}$ degrees of freedom. The quantity ${D^{(n)}_{\rm MD}}^2$ can also be decomposed into components, e.g., via a pivoted Cholesky decomposition of the matrix $K$. This decomposition can be used to extract meaningful information about whether each curve $c_n(q_i)$ is described by the hypothesized GP. These tests are implemented in the python package {\tt gsum}~\cite{gsum}, are described in Section~III of Ref.~\cite{Millican:2024yuz}, and are derived and explained in detail in Ref.~\cite{BastosDiagnosticsGaussianProcess2009}.

The curves $c_n$ will best fulfill the criterion that they are described by a GP for a particular value of the $\chi$EFT breakdown scale $\Lambda_b$.  The most probable $\Lambda_b$ is the value that yields a value of $Q$ for which the set of curves $c_n$ are all closest in size, i.e., are described by a common variance $\bar{c}^2$. 
The outputs of a GP diagnostics analysis are, therefore, a breakdown scale $\Lambda_b$ that maximizes the probability that the coefficient curves $c_n$ are all draws from a single GP, together with the correlation length $\ell_q$ and variance $\bar{c}^2$ that describes those curves. 
Formulae for the joint probability distribution function of $\ell_q$ and $\Lambda_b$ can be found in the Appendix of Ref.~\cite{Melendez:2019izc} and are implemented in the {\tt gsum} package.

\begin{figure}[hbt]
\includegraphics[width=0.48\linewidth]{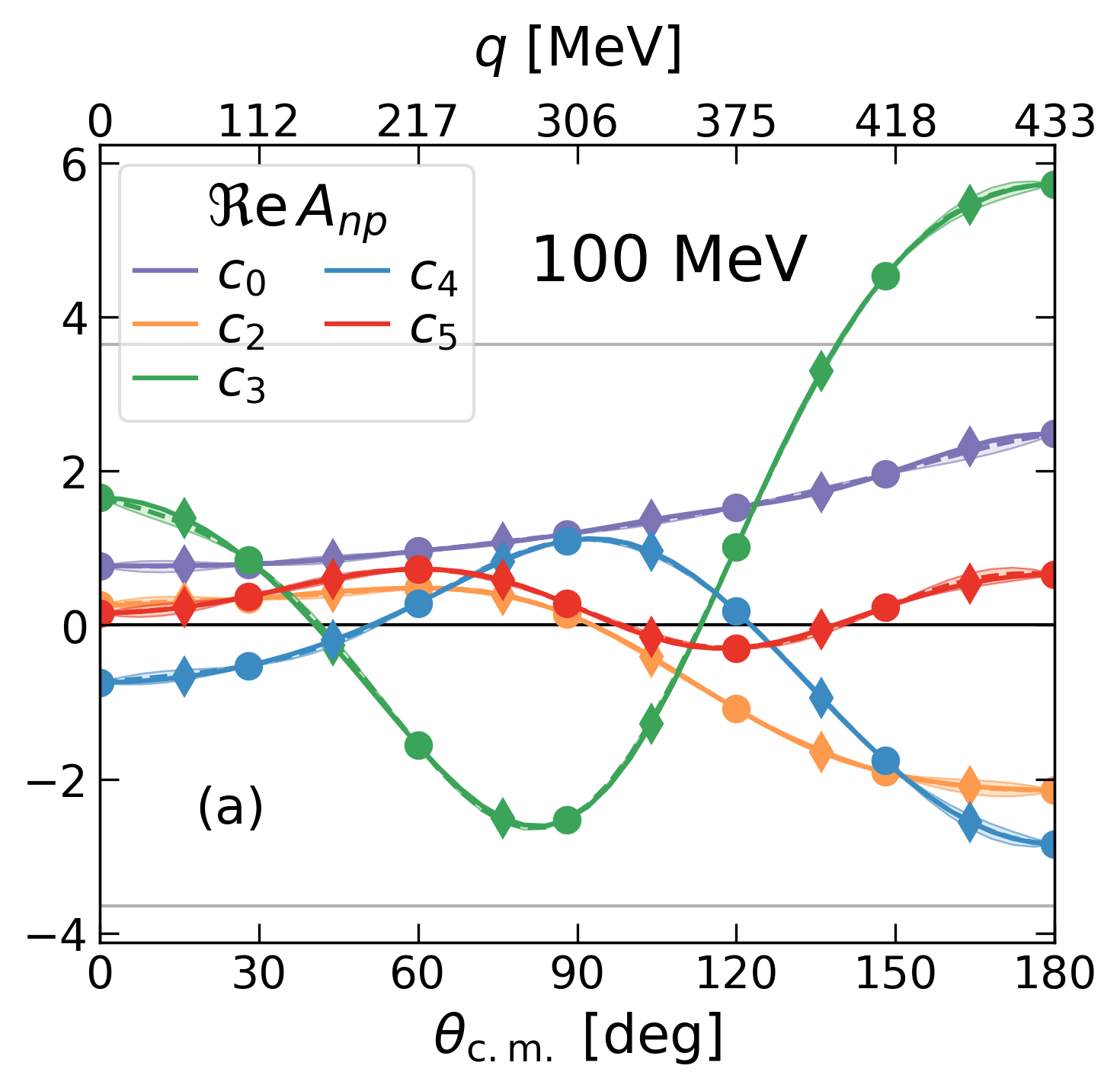}
\includegraphics[width=0.48\linewidth]{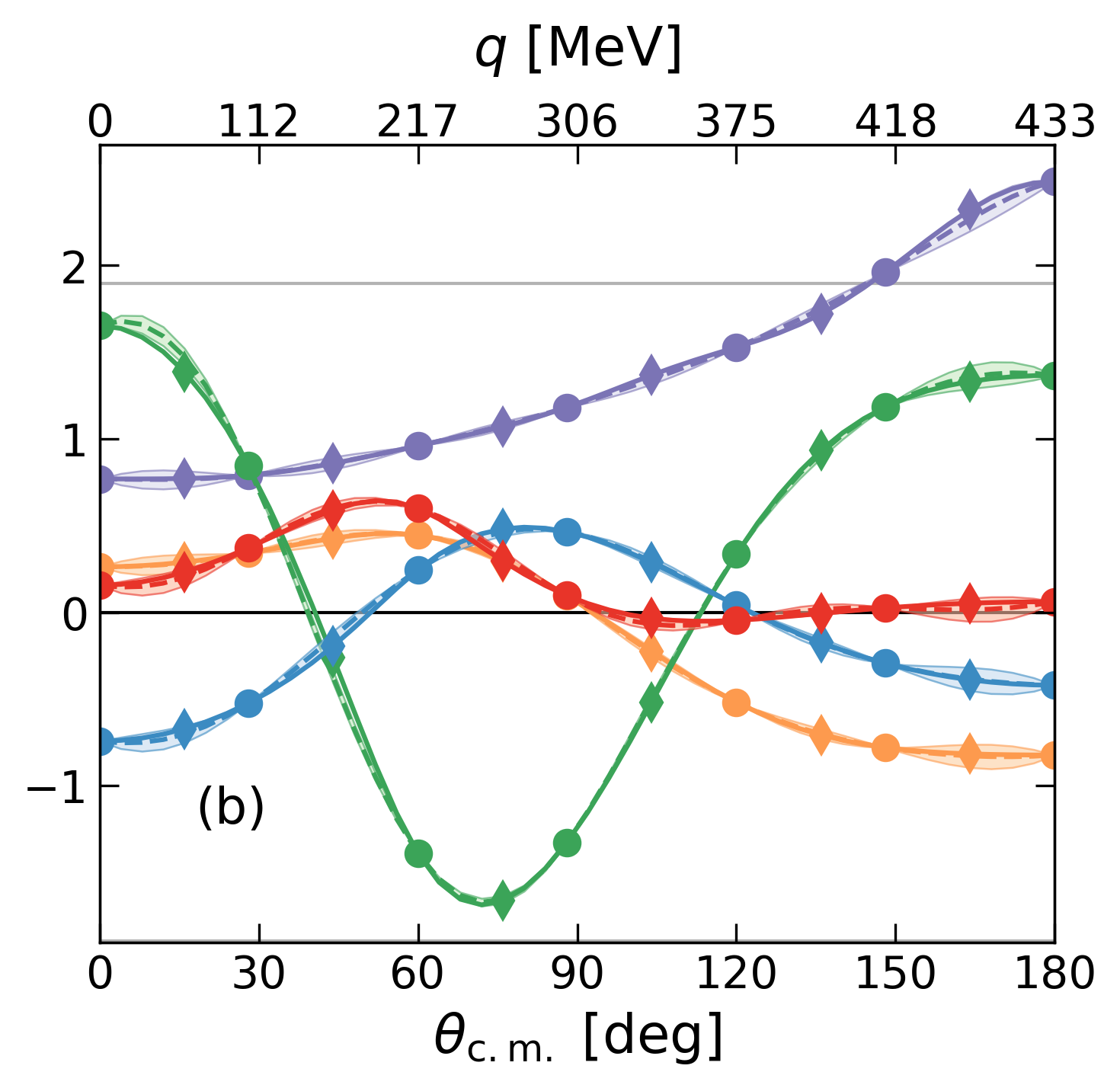}
\caption{The coefficient curves for \Renp{A} at $\elab$ = 100~MeV with $Q$ being independent of $\theta_{\rm{c.m.}}$ (a) and dependent on $\theta_{\rm{c.m.}}$ (b).
The (barely discernible) bands around the GP emulated curves (dashed) represent 2$\sigma$ uncertainty. The filled points are the training points, while the filled diamonds are the testing points. The gray horizontal lines indicate plus and minus twice the fitted value for $\bar{c}$. Compared to (a), the sizes of the coefficient curves in (b) are more similar and natural-sized across the different orders. Additionally, the sizes of the curves do not significantly grow or shrink as $\theta_{\rm c.m.}$ increases.}
\label{fig2}
\end{figure}

The main goal of this procedure is to generate statistically principled uncertainties for the $\chi$EFT prediction of the Wolfenstein amplitudes at order $k$. The Credible Interval Diagnostic $D_{\rm CI}$, also implemented in the package {\tt gsum}, tests whether this goal has been achieved. 
$D_{\rm CI}$ assesses whether the credibility intervals assigned at each order cover the corresponding percentage of points in a particular Wolfenstein amplitude. 

This is not the same as checking the consistency of the EFT error bars with the order-by-order convergence of the amplitude, but it does check the consistency of those error bars with ``data". In this case, we take the Wolfenstein amplitudes constructed from the high-precision CD-Bonn potential~\cite{Machleidt:2000ge} as a proxy for experimental data since that potential fits the NN database available when it was constructed in 2001 with a $\chi^2$ per degree of freedom of 1.01 for the $pp$ data and 1.02 for the $np$ data.

\subsection{Specific analysis choices}
\label{subsec:choices}

\begin{figure}[hbt]
\includegraphics[width=0.2\linewidth]{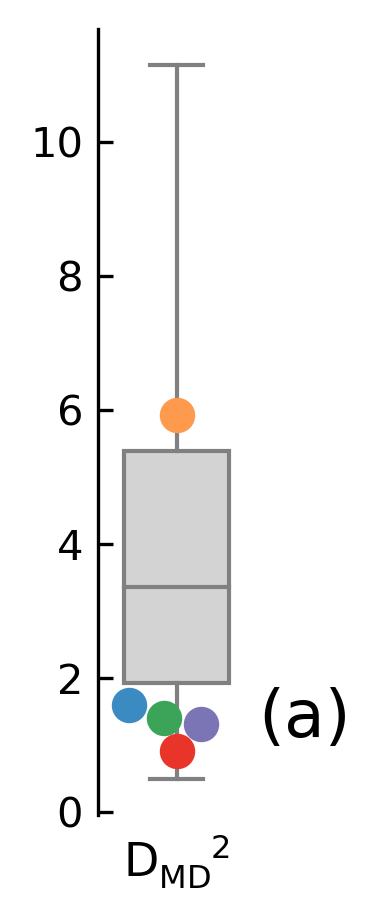}
\hspace{4mm}
\includegraphics[width=0.251\linewidth]{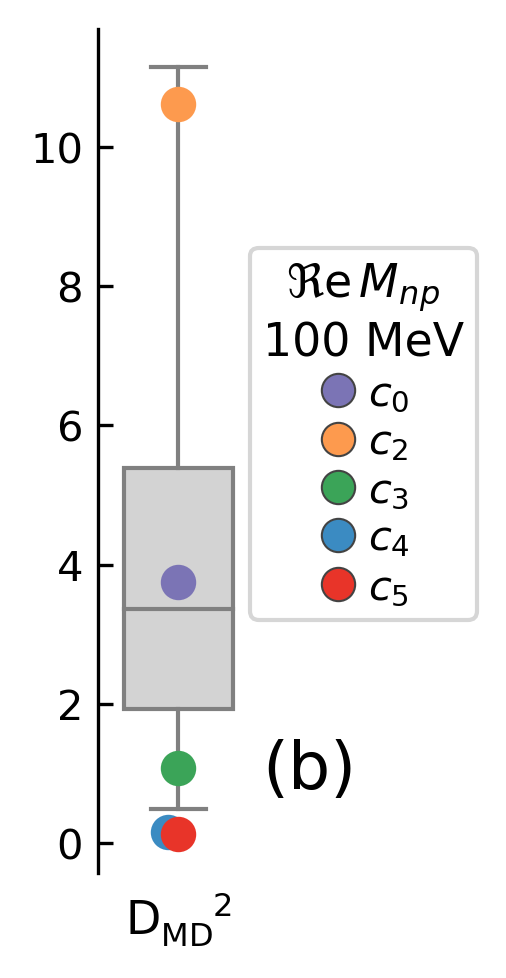}
\caption{The square of the Mahalanobis distances for \Renp{M} at $\elab = 100$~MeV using $Q$ independent of momentum transfer with $\yref = 1.0$~fm (a) and Q dependent on momentum transfer with $\yref = \Im$m$\,A_{np}$ at 5th order (b). Compared to (a), the spread of squared Mahalanobis distances follows a $\chi^2$ distribution better. The horizontal lines, shaded box, and error bars on the Mahalanobis distance plot indicate the expected median, 50\% credible interval, and 95\% credible interval. 
}
\label{fig3}
\end{figure}
We now turn to the specific choices we make in analyzing the Wolfenstein amplitudes, defined in Subsection~\ref{subsec:Wolfensteins}, using the Gaussian process methodology first developed by Melendez et al.~\cite{Melendez:2019izc}, and described in Subsection~\ref{subsec:GPs}. 

Figures~\ref{fig2} and \ref{fig3} present curves and diagnostics supporting these choices for the case where coefficient curve data is treated as a function of angle. This is the input space in which the analysis of Ref.~\cite{Millican:2024yuz} was performed. The formulae given in Subsection~\ref{subsec:GPs} apply in this input space of c.m. scattering angle, with the simple replacement of $q$ by $\theta_{\rm c.m.}$, and the mapping
$q = 2p\sin(\theta_{\rm c.m.}/2)$.
However, treating the coefficient curves as a function of the momentum transfer, $q$ (cf. the discussion at the end of Sec.~\ref{sec:pn}) is more natural. The improvements noted in Figs.~\ref{fig2} and \ref{fig3} are not altered by this different choice of input space.

This work does not treat the Wolfenstein amplitudes as GPs in a two-dimensional input space of relative momentum, $p$, and momentum transfer, $q$. Instead, we choose five energies and perform the one-dimensional analysis described above. The laboratory energies we select, 25 MeV, 50 MeV, 100 MeV, 150 MeV, and 200 MeV, range from one that is well within the domain of validity of pionless EFT~\cite{Ekstrom:2024dqr} to one where the expansion parameter of $\chi$EFT is of order 0.5.

When we report our final results, we decline to combine the PDFs obtained for random variables across different energies since it is not clear that the five sets of results are fully independent. However, the energies are far enough apart that the analysis has different features at each one. And the difference of roughly 50 MeV in center-of-mass momentum between each one is only slightly less than the momentum correlation length  $\ell_{p} \approx 60$ MeV, found by Millican et al. in a two-dimensional GP analysis of NN observables~\cite{Millican:2024b}.

As discussed extensively in Millican et al.~\cite{Millican:2024yuz}, the output of the GP analysis depends on the choice of expansion parameter. Here we take 
\begin{equation}
    Q=\frac{{\rm smax}(p,q) + m_\pi}{\Lambda_b + m_\pi},
    \label{eq:Qchoice}
\end{equation}
where ${\rm smax}$ is a smooth maximum of the relative momentum $p$ and the momentum-transfer $q$, $m_\pi$ is the pion mass, and $\Lambda_b$ is the $\chi$EFT breakdown scale. The relative momentum is given by
\begin{equation}\label{eq:prel}
    p^2 = \frac{m_2\pow{4} \elab (\elab + 2m_1)}{(m_1 + m_2)^2 + 2m_2 \elab},
\end{equation}
where $m_1$ is the projectile mass, $m_2$ is the target mass, and $\elab$ is the laboratory kinetic energy in the lab frame. The smooth maximum we use follows Eq. (12) in Ref.~\cite{Millican:2024yuz},
\begin{equation}\label{eq:smax}
    {\rm smax}(p,q) = \frac{1}{N} \log_{1.01}(1.01^{Np} + 1.01^{Nq})\,,
\end{equation}
with $N = 5$.
Our choice of expansion parameter differs from Millican et al.~\cite{Millican:2024yuz} in two ways. First, we include the momentum transfer $q$ as one of the soft scales that can enter the expansion parameter. Second, we do not vary the explicit chiral-symmetry-breaking scale $m_\pi$ to optimize the expansion parameter, instead keeping it at its physical value (cf. Ref.~\cite{Epelbaum:2019wvf} for arguments why it should be varied). The effects of $Q$ being either independent or dependent on the momentum transfer, along with an example set of training and testing points, can be seen in Fig.~\ref{fig2}. For an expansion parameter independent of momentum transfer (left panel), the curves for $c_4$ and especially $c_3$ become quite large at backward angles, which does not agree well with the BUQEYE model. Upon employing a $q$-dependent expansion parameter (right panel), the sizes of the coefficient curves become more similar and natural-sized across the different orders. Furthermore, only $c_0$ goes outside the $2\sigma$-band at backward angles, and the other curves do not significantly grow or shrink with increasing scattering angle.

Several factors constrain the number of training points. More training points result in a more accurate GP emulator. However, too many points can lead to underestimated correlation lengths, truncation errors, or, at worst, a singular covariance matrix. Too few training points can result in a poor GP estimate. In particular, the GP may revert to its mean of zero, and the marginal variance between training points may become large if they are too sparsely spaced.
 For these reasons, we choose the minimum possible number of training points that do not lead to such issues with the GP emulation. Meanwhile, the testing points are selected to be as close to the middle between training points as possible on a discrete grid. Having too few testing points does not provide a stringent statistical test of the GP, while too many can produce artificially large Mahalanobis distances. This procedure is carried out for each energy and amplitude we analyze.

For $y_{\rm ref}$, the overall scale of the observable, we choose the imaginary part of the $np$ Wolfenstein amplitude $A$ at the energy being analyzed, as computed up to N$^4$LO in $\chi$EFT. Here, it should be noted that we took the freedom to scale \Imnp{A} and, in some cases applied an offset.  We use this $y_{\rm ref}$ for both the $np$ and $pp$ amplitudes. Other choices were examined, but either they did not capture the overall features of the momentum-transfer dependence of all the amplitudes, or they had other angular structures that, upon division, created artificial effects in the coefficient curves for amplitudes other than $A$ itself.

The use of \Imnp{A} for $\yref$ and Eq.~\eqref{eq:Qchoice} for $Q$ is supported by overall improvements in the Mahalanobis distance and credible interval diagnostics (across all amplitudes). Specifically, the spread of squared Mahalanobis distances better follow a $\chi^2$ distribution (see Fig.~\ref{fig3} for an example), and the credible intervals are more consistent with the amplitudes given by the CD-Bonn potential.



\section{Neutron-Proton Amplitudes}
\label{sec:pn}

 In this Section, we discuss the analysis of the Wolfenstein amplitudes for $np$ scattering from 25 to 200 MeV laboratory kinetic energy. The guide to choices for the analysis is 
 justified in Subsection~\ref{subsec:choices}. The energy of 100~MeV is sufficiently high for partial waves up to about $J=3$ to contribute (where $J$ is the total angular momentum quantum number), meaning that features we encounter at 100 MeV will likely also be present at higher energies.  
We first discuss what those analysis choices produce for the coefficient curves at 100 MeV laboratory kinetic energy. We then show that insights seen there also hold at 25 and 200 MeV. (We also checked the behavior of the coefficient curves at 50 and 150 MeV, but since it was consistent with the other energies, we do not display those results here.)

\begin{figure}[hbt]
\includegraphics[width=0.75\prcColumnWidth]{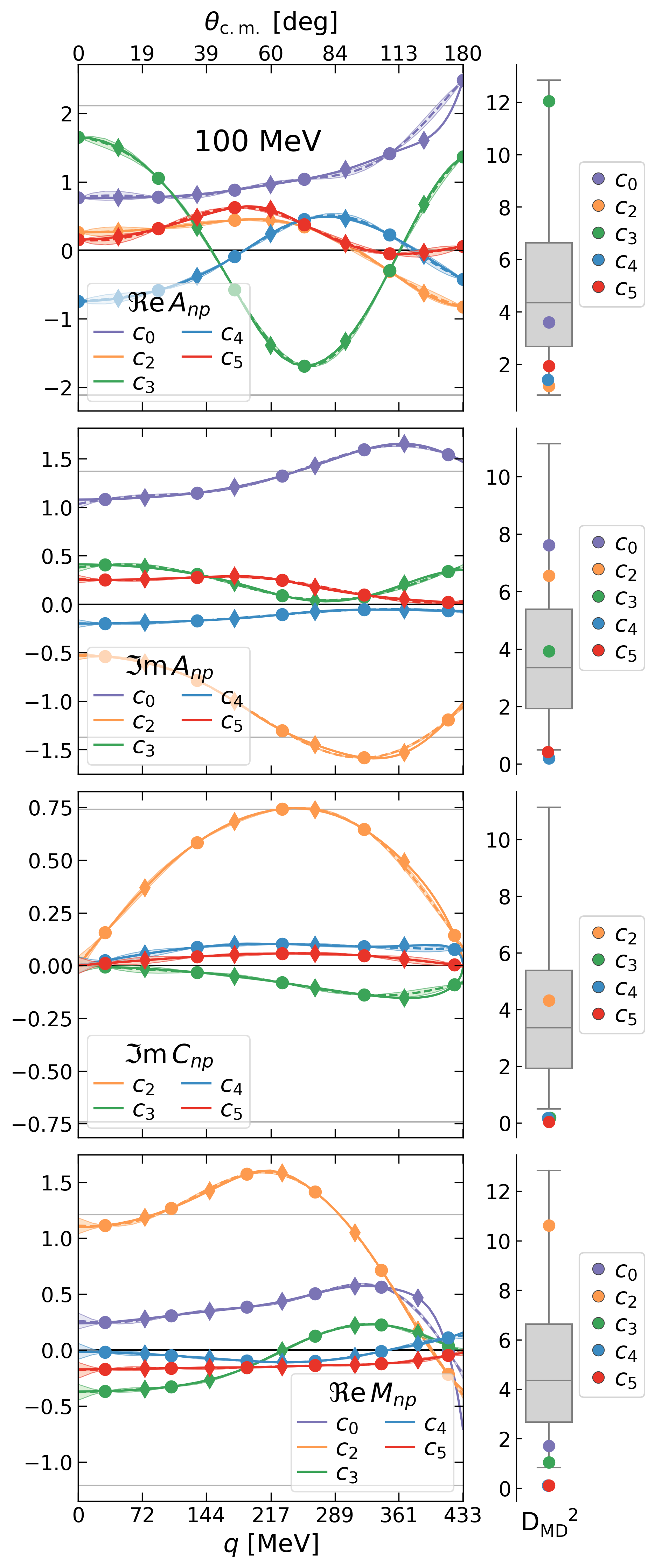}
\caption{Coefficient curves and Mahalanobis distance diagnostics for selected $np$ Wolfenstein amplitudes at 100~MeV. The panels showing the coefficient curves have the same legend as in Fig.~\ref{fig2}, and the Mahalanobis distances are depicted as in Fig.~\ref{fig3}. The gray horizontal lines indicate plus and minus the twice the fitted value for $\bar c$.
}
\label{fig4}
\end{figure}

\begin{figure}[hbt]
\centering
\includegraphics[width=0.49\prcColumnWidth]{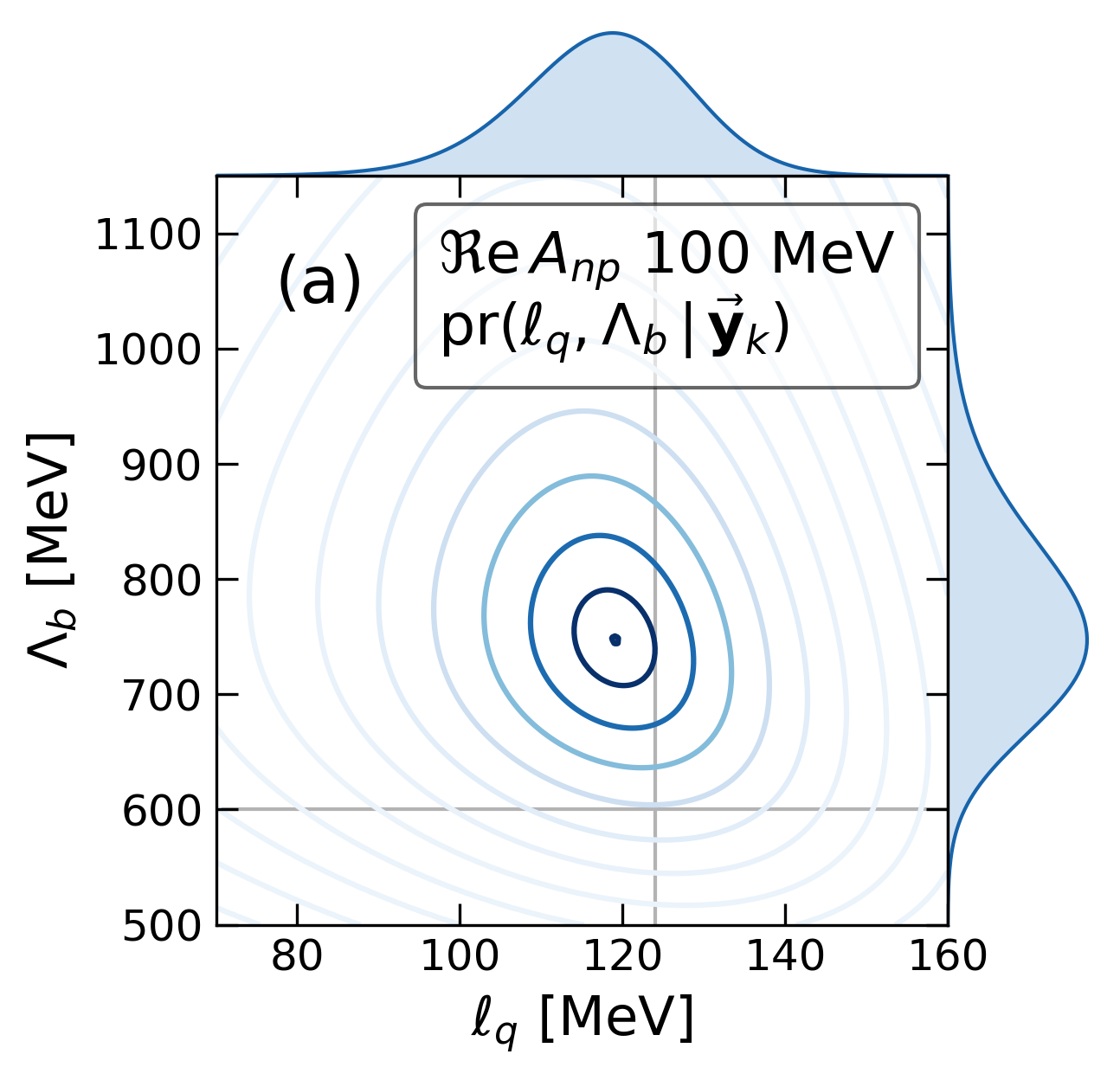}
\includegraphics[width=0.49\prcColumnWidth]{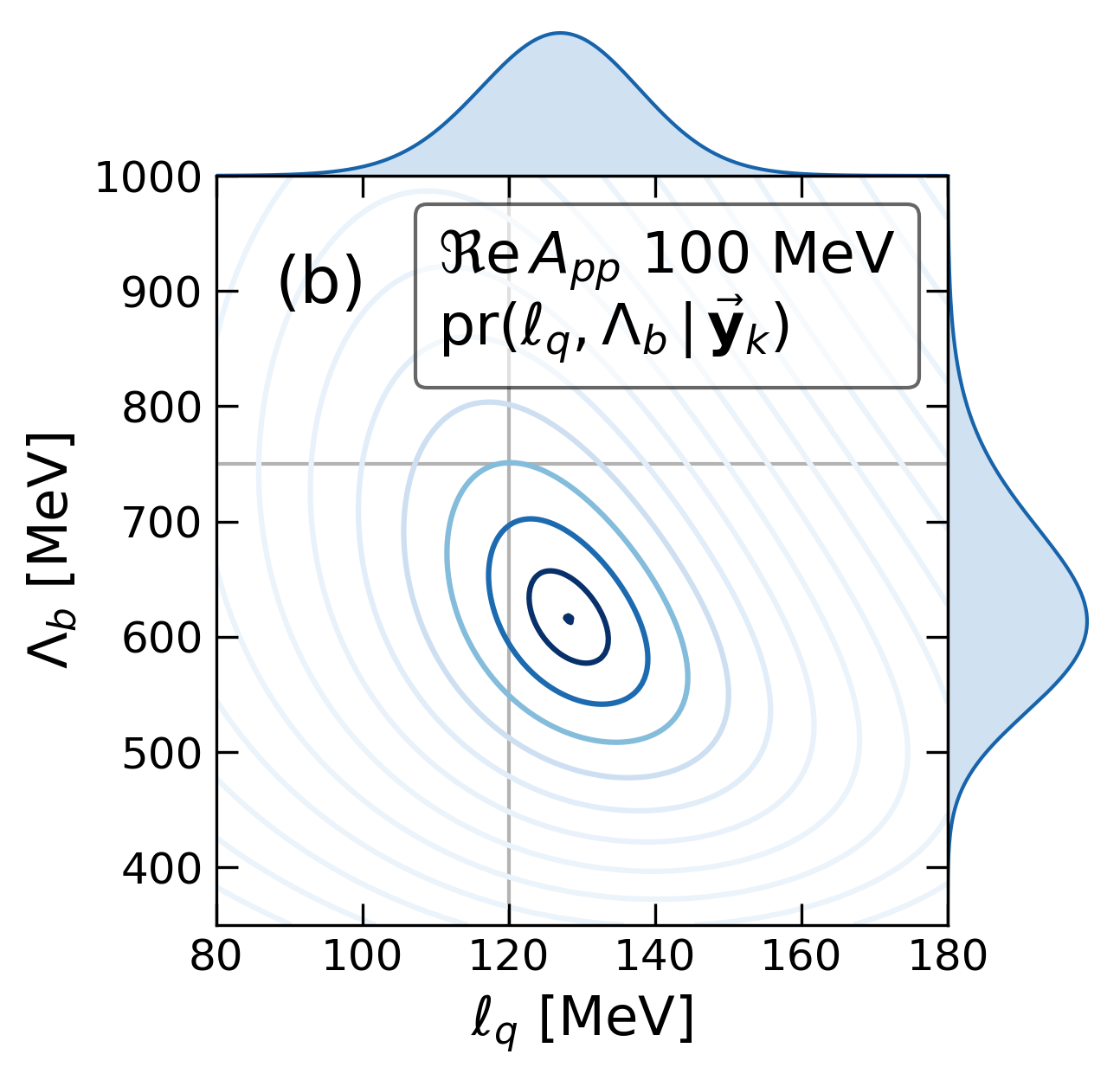}
\caption{Joint $\ell_q-\Lambda_b$ posteriors extracted from \Renp{A} (left panel) and \Repp{A} (right panel) at 100~MeV with a uniform prior $pr(\ell_q,\Lambda_b) \propto 1$. The contours increment in half-standard deviation intervals, with the dark blue point denoting the maximum {\it a posteriori} value. The
one-dimensional distributions for the breakdown scale $\Lambda_b$ and the correlation length $\ell_q$ are shown at the axes.
}
\label{fig5}
\end{figure}
\begin{figure}[h!]
\includegraphics[width=0.7\prcColumnWidth]{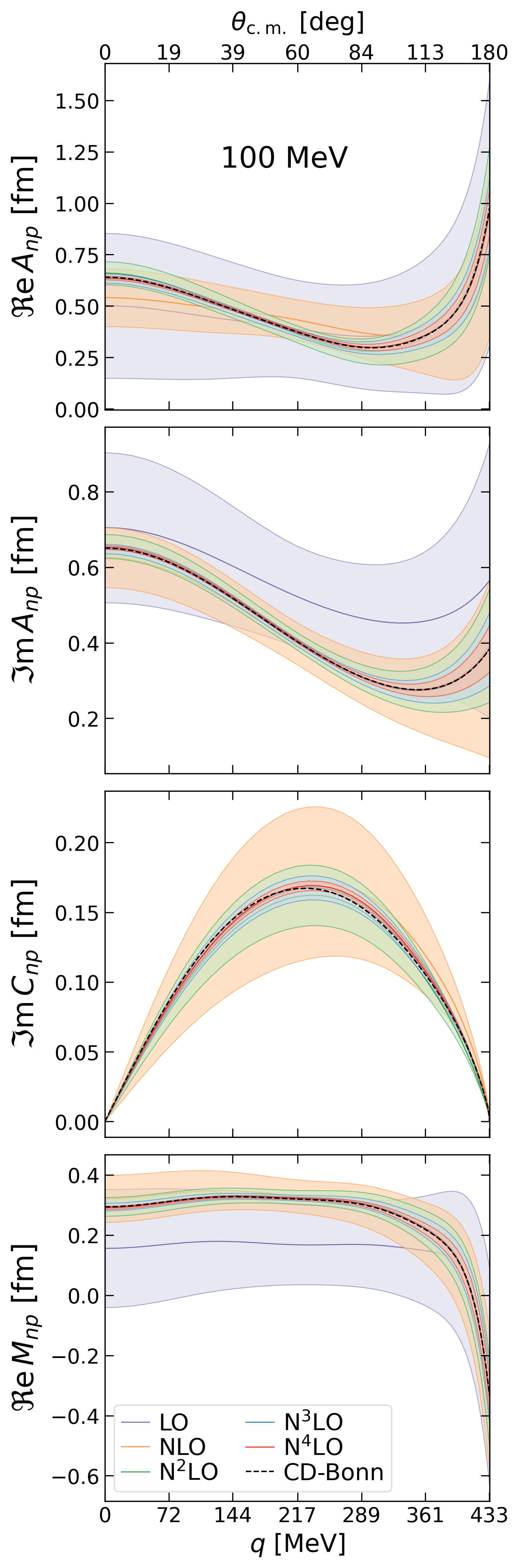}
\caption{The predictions and their corresponding 2$\sigma$ truncation error estimates obtained with different $\chi$EFT orders of the SCS interaction for $np$ scattering Wolfenstein amplitudes at 100~MeV. These are generated via the process depicted in Fig.~\ref{fig4}. The black dashed line results from a calculation based on the CD-Bonn potential. 
}
\label{fig6}
\end{figure}

After establishing the choices of $Q$ and the training and testing points according to the principles in subsection~\ref{subsec:choices}, we analyze the other Wolfenstein amplitudes at 100~MeV laboratory kinetic energy. In Fig.~4, we show the coefficient curves and corresponding Mahalanobis distances for the real and imaginary parts of $A_{np}$, the imaginary part of $C_{np}$, and the real part of $M_{np}$.
These amplitudes are reasonably large, as seen in Fig.~\ref{fig1}, and reflect the scalar, vector, and tensor operator structure of the interaction. We also note that they are the amplitudes contributing to nucleon-nucleus scattering from nuclei with $0^+$ ground states in an {\it ab initio} framework~\cite{Burrows:2020qvu}.

We observe that the coefficient curves for $c_0$ are relatively large at backward angles in \Renp{A} and \Imnp{A}. For both, the charged one-pion exchange in LO is most likely responsible. Coefficient curves for \Renp{G} and \Renp{H} are given in the Appendix~\ref{appendixa} and show similar characteristics.
However, the coefficient curve $c_2$ is larger than the other coefficients in each of the second to fourth panels of Fig.~\ref{fig4}. For \Imnp{A}, \Imnp{C}, and \Renp{M} it even exceeds 
 2${\bar c}$ in certain regions of momentum space. This is also reflected in the size of the corresponding Mahalanobis distances.
We reiterate that, for the case of \Imnp{C}, we are looking at a vector (spin-orbit) operator. This operator has no LO contribution in the standard $\chi$EFT power counting. Thus, $c_2$ is {\it de facto} the ``leading" contribution to this Wolfenstein amplitude. 
Similarly, the coefficient curves $c_2$ for \Imnp{A} and  \Renp{M} originate from the relatively large difference between the LO and NLO contributions in Fig.~\ref{fig1},  so the rather discrepant $c_2$ revealed by
{\tt gsum}'s diagnostic tools are not very surprising.  That
$c_2$ is significantly larger than other coefficients in several amplitudes
may indicate that the power counting in this specific SCS $\chi$EFT NN interaction is not optimal. Nogga {\it et al.} have argued on renormalization-group grounds that certain NLO terms in this interaction should be promoted to LO, which might improve the convergence pattern in this type of analysis~\cite{Nogga:2005hy}.

In Fig.~\ref{fig5}, panel (a), we show the joint posterior obtained for the breakdown scale $\Lambda_b$ and the correlation length $\ell_q$ from our analysis of \Renp{A}. We used a uniform prior $pr(\ell_q,\Lambda_b) \propto 1$.
The blue point is the maximum {\it a posteriori} (MAP) value, and the contours indicate
probability increments corresponding to half of a standard deviation.
From this, we can deduce that the estimated breakdown scale is about 750~MeV, close to the mass of the $\rho$-meson. The estimated momentum correlation length is about 120 MeV. 

Henceforth, we adopt the value $\Lambda_b=750$~MeV for our analysis since this is close to the maximum {\it a posteriori} value for $\Lambda_b$ from Fig.~\ref{fig5} and is also supported by analogous results from other amplitudes presented at the end of this section. 
Figure~\ref{fig6} shows the $\chi$EFT predictions at 100 MeV and their corresponding 2$\sigma$ uncertainties at different orders of the SCS interaction. As already indicated by the fact that $c_2 > 2 \bar{c}$,
the LO truncation error band does not cover NLO (and higher-order) results for \Imnp{A} and \Renp{M}. For these two amplitudes, it is only at NLO, and beyond, that the uncertainty at N$^k$LO encompasses the N$^{k+1}$LO result. In fact, the truncation errors fulfill this expectation for \Renp{A} starting only at N$^2$LO.
Meanwhile, the imaginary part of $C_{np}$ shows complete coverage of the $\chi$EFT truncation errors. However, as mentioned earlier, this amplitude has no LO contribution since the spin-orbit contribution only enters at NLO. Hence, we only construct truncation errors for \Imnp{C} starting at NLO.

The 95\% $\chi$EFT error bands for \Renp{A}, \Imnp{A}, and \Renp{M} all encompass the CD-Bonn results. For \Imnp{C}, a tiny tension exists between $\chi$EFT and CD-Bonn at forward and intermediate angles. 

\begin{figure}[h!]
\includegraphics[width=0.7\prcColumnWidth]{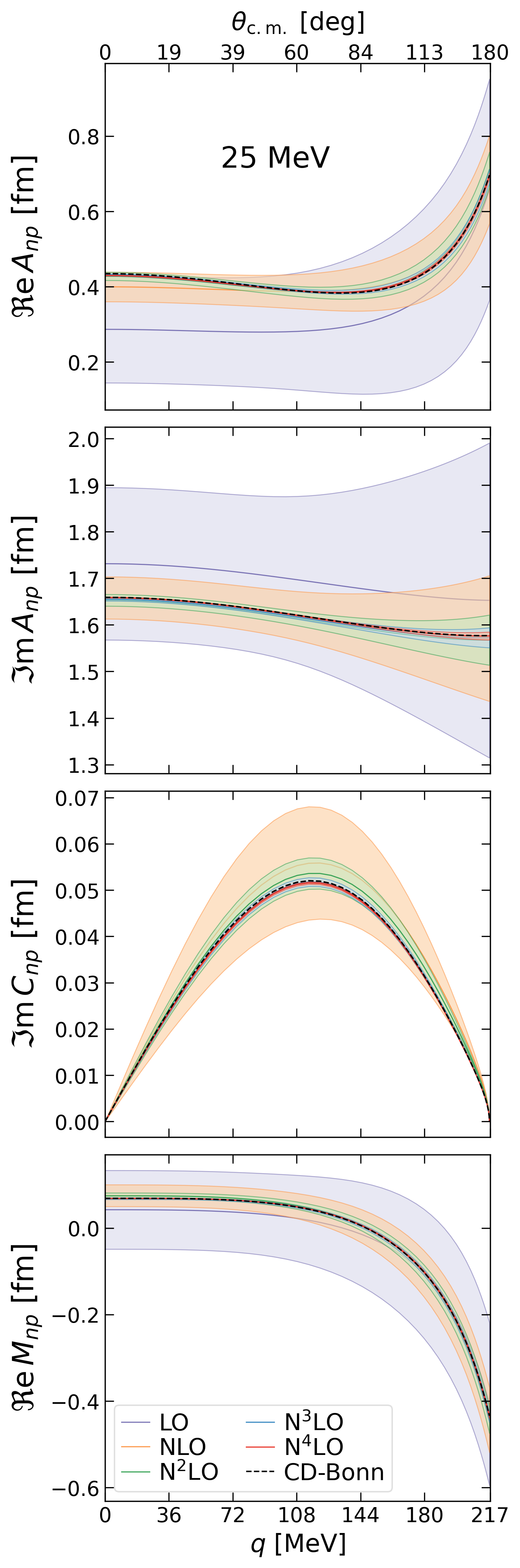}
\caption{The predictions and their corresponding 2$\sigma$ truncation error estimates obtained with different $\chi$EFT orders of the SCS interaction for $np$ scattering Wolfenstein amplitudes at 25~MeV. These are generated via a process similar to the one depicted in Fig.~\ref{fig4}. The black dashed line results from a calculation based on the CD-Bonn potential.
}
\label{fig7}
\end{figure}

\begin{figure}[b!]
\includegraphics[width=0.7\prcColumnWidth]{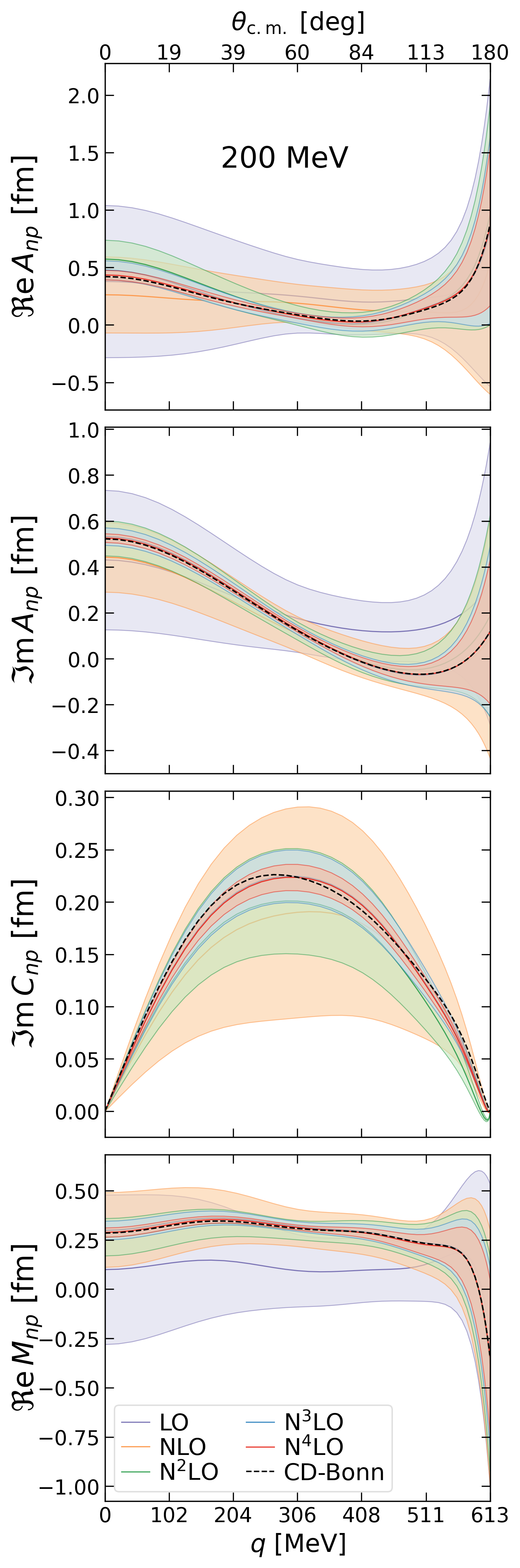}
\caption{The predictions and their corresponding 2$\sigma$ truncation error estimates obtained with different $\chi$EFT orders of the SCS interaction for $np$ scattering Wolfenstein amplitudes at 200~MeV. These are generated via a process similar to the one depicted in Fig.~\ref{fig4}. The black dashed line results from a calculation based on the CD-Bonn potential.
}
\label{fig8}
\end{figure}

We conducted the same full analysis for $np$ Wolfenstein amplitudes at lower and higher energies. This necessitated the determination of training and testing points, as well as checking coefficient curves and Mahalanobis distances. Since this procedure gave overall results similar to the ones shown in Fig.~\ref{fig4}, we do not show them here. To contrast our findings with the results at 100~MeV, we show in Fig.~\ref{fig7} the predictions and corresponding truncation errors for 25~MeV, and in Fig.~\ref{fig8} results for 200~MeV.

The scattering at lower energies is mainly governed by S- and P-wave scattering. The Wolfenstein amplitudes at 25~MeV receive most of their contribution from these partial waves. To assess the size of P-wave contributions, we compare the magnitudes of \Imnp{C}, the spin-orbit contribution to the NN interaction, at 25~MeV and 100~MeV. \Imnp{C} at 25~MeV is more than a factor of two smaller than at 100~MeV. It also converges very quickly with increasing chiral order. For central and tensor contributions, the leading $\chi$EFT order describes the S-waves, which dominate at 25~MeV, moderately well. 
In contrast to the analysis at 100~MeV, the truncation error bands for \Imnp{A} and \Renp{M} follow a pattern where the respective lower order entirely encloses the error bands for each additional higher order, and the expansion converges very quickly. Moreover, the very small N$^4$LO error bands again encompass the ``experimental'' amplitudes reconstructed using the high-precision CD-Bonn potential.  The only exception is \Renp{A}, where, at small angles, the LO and NLO error bands barely cover the error bands of the higher orders. 

At 200~MeV, the highest laboratory kinetic energy we analyze, the situation is similar to the results at 100~MeV. But it also shows some notable differences. The predictions of the amplitudes and their corresponding 2$\sigma$ truncation error estimates for the different $\chi$EFT orders are shown for this energy in Fig.~\ref{fig8}.  

Our calculations sum partial waves up to $J=6$. However, partial waves with $l$=1, 2, and 3 mainly contribute to the amplitudes. At this energy, the contribution of the two S-waves compared to P- and D-waves is small since both S-wave phase shifts come close to their zero-crossing. 

We note that, for \Renp{A}, the backward angles (higher momenta) show a relatively steep rise. In addition, the expansion parameter becomes large in this region, greater than 0.9. The error bars then become large for all orders. 
For both \Renp{A} and \Imnp{A}, there are large changes (50-100\%) in these amplitudes from LO to the final result in this momentum region. The (large) 95\% LO error bar manages to cover the N$^4$LO and CD-Bonn results in most of the angular range but does not do so at intermediate angles for \Imnp{A}. For \Renp{M}, which also exhibits a steep amplitude change at large momentum transfers, the situation looks more systematic in that region in the sense that truncation errors are all located within the LO error bar. However, at intermediate angles, the LO curve is again more than 2$\sigma$ away from the N$^4$LO and CD-Bonn results---as we also saw for the error estimates at 100~MeV. Comparing this to the amplitude itself calculated at different orders of the SCS interaction, Fig.~\ref{fig1}, we see again that this is a region where the leading order of the SCS interaction is still far away from the final amplitude. So, the estimated LO truncation error is too small there. 

Turning our attention to \Imnp{C}, it is worthwhile pointing out that, with increasing laboratory energy, the size of the amplitude, driven by the P-waves, increases, and so do the truncation errors.  $\chi$EFT is internally consistent in this amplitude at NLO and beyond. However, the shape of the CD-Bonn \Imnp{C} is noticeably different at $E_{\rm lab}=200$ MeV, leading to some tension with even the N$^4$LO uncertainty band at very forward and very backward angles.

\begin{figure}[hbt]
 \includegraphics[width=0.59\linewidth]{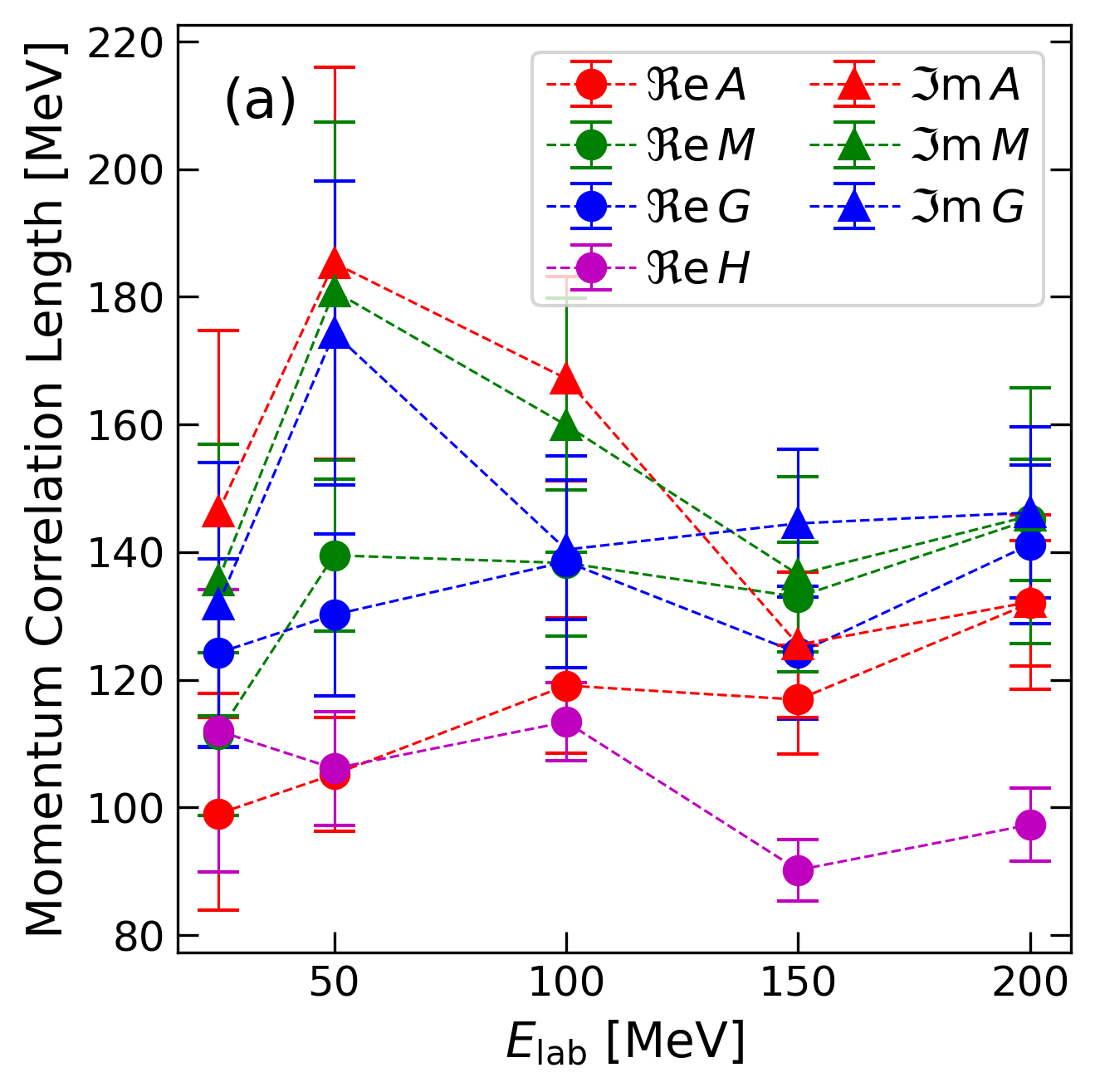}
 \includegraphics[width=0.59\linewidth]{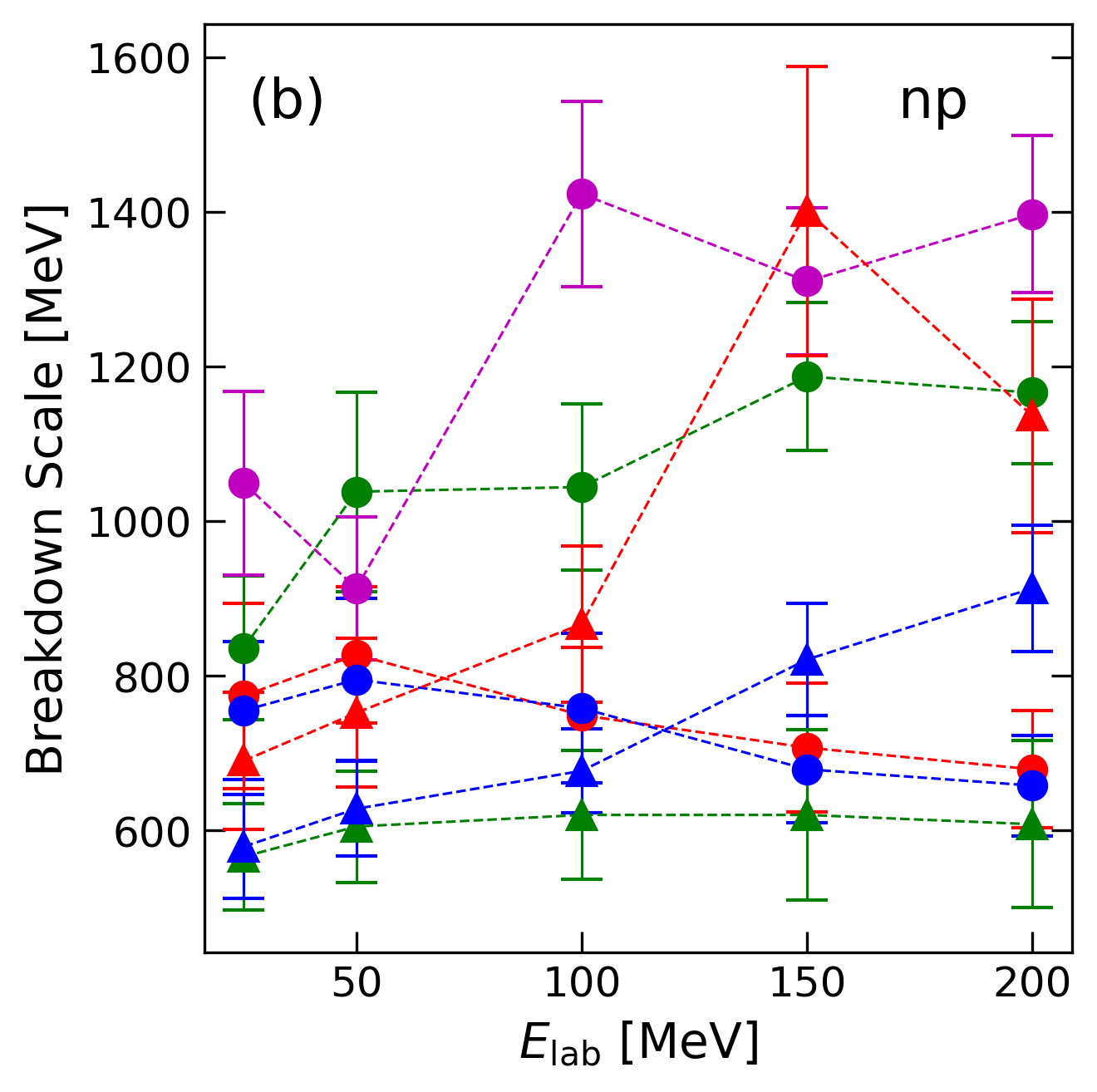}
\caption{Maximum {\it a posteriori} values for momentum correlation lengths, $\ell_q$ and breakdown scales, $\Lambda_b$ for selected $np$ Wolfenstein amplitudes with 1$\sigma$ error bars. The momentum correlation lengths are approximately constant within error bars across energy. 
}
\label{fig9}
\end{figure}

For each energy and amplitude, we generated a two-dimensional posterior similar to the one shown in Fig.~\ref{fig5} for \Renp{A} and \Repp{A} at 100 MeV. Figure~\ref{fig9} summarizes the 68\% intervals for $\ell_q$ and $\Lambda_b$ obtained for seven different amplitudes at each of the five different energies for which we carry out our analysis. Note that \Imnp{C} is not included here for the reasons already discussed, and \Renp{C} and \Imnp{H} are omitted from consideration because they are small. 

In the real part of a given amplitude, the correlation length in $q$ is constant across the different energies within uncertainty. This finding is important since a roughly constant momentum correlation length for the operators building the chiral EFT can not {\it a priori} be expected. It is also interesting to note that this momentum correlation length corresponds approximately to the mass of the pion.  During our work, we also studied the correlation as a function of the scattering angle. For that, we found that the angular correlation length decreases as a function of the relative momentum, $p$, roughly as $1/p$. 
From this finding we concluded that a momentum correlation length is the better quantity to consider. A similar result regarding the behavior of the angular correlation length was also found in Ref.~\cite{Millican:2024yuz}.

The breakdown scale $\Lambda_b$ is also quite stable across energies for \Imnp{M}, \Renp{A}, and \Renp{G}. None of the analyses suggest a breakdown scale lower than 600 MeV, and several give breakdown scales markedly higher. The average maximum {\it a posteriori} value for the breakdown scale is 860~MeV. However, \Renp{H}'s $\Lambda_b$ varies significantly with energy and is also higher than that obtained from most amplitudes. Without these data included, the average $\Lambda_b$ is 800 MeV. We do caution that averaging across amplitudes and energies is not necessarily a statistically sound procedure here since $\Lambda_b$'s found at nearby energies will undoubtedly be correlated. There may also be correlations between the $\Lambda_b$'s found in different Wolfenstien amplitudes.


\section{Proton-Proton Amplitudes}
\label{sec:pp}

In this Section, we discuss the analysis of the Wolfenstein amplitudes for $pp$ scattering in the energy range from 25 to 200 MeV laboratory kinetic energy. We employ the same choices as given in 
Subsection~\ref{subsec:choices}, and 
proceed similarly to the treatment of $np$ scattering in the previous section. First, we examine the order-by-order $\chi$EFT coefficient curves for several Wolfenstein amplitudes at 100~MeV laboratory kinetic energy. Then, we discuss the $\chi$EFT uncertainties derived from those curves at that energy. Finally, we highlight the similarities and differences in the Wolfenstein amplitudes and their $\chi$EFT uncertainties at 25 and 200 MeV. Throughout this discussion, we use $\Lambda_b=750$ MeV.

\begin{figure}[hbt]
\includegraphics[width=0.75\prcColumnWidth]{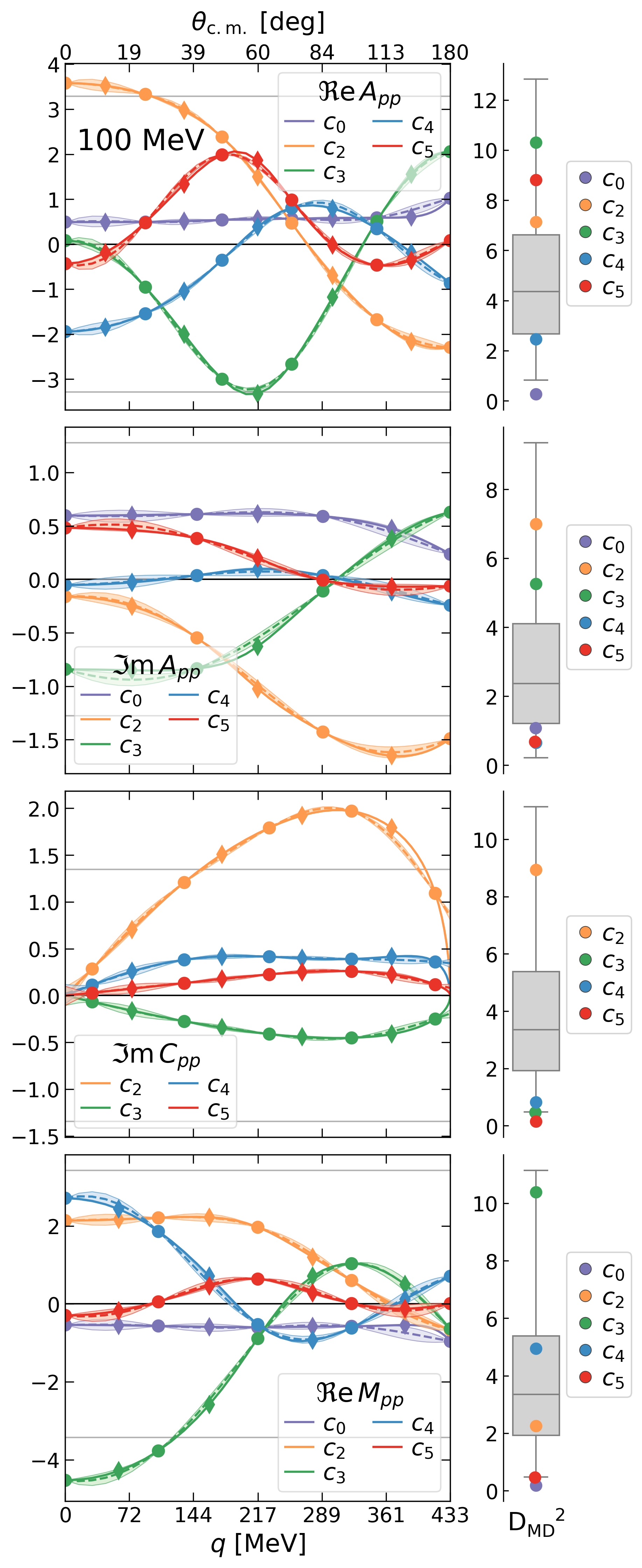}
\caption{Coefficient curves and Mahalanobis distance diagnostics for selected $pp$ Wolfenstein amplitudes at 100~MeV. The gray horizontal lines indicate plus and minus the expected value for 2$\bar{c}$.}
\label{fig10}
\end{figure}

\begin{figure}[b!]
\includegraphics[width=0.7\prcColumnWidth]{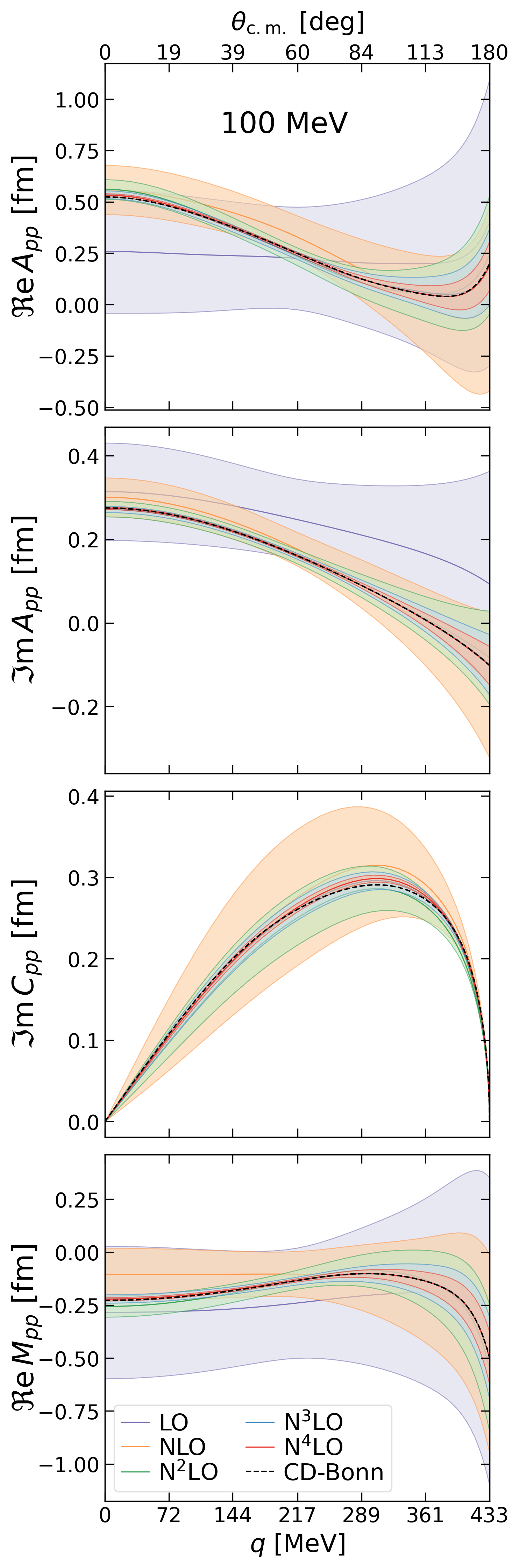}
\caption{The predictions and their corresponding 2$\sigma$ truncation error estimates obtained with different $\chi$EFT orders of the SCS interaction for $pp$ scattering Wolfenstein amplitudes at 100~MeV. These are generated via the process depicted in Fig.~\ref{fig10}. The black dashed line results from a calculation based on the CD-Bonn potential.
}
\label{fig11}
\end{figure}

Since in $pp$ scattering only the isospin-1 channels of the NN interaction contribute, there are fewer partial waves in the sum for a particular Wolfenstein amplitude. We do not include the Coulomb force in the $pp$ amplitudes, thus only considering the nuclear `bar' phase shifts~\cite{Stapp:1956mz} in the partial-wave summation for obtaining the Wolfenstein amplitudes.

The coefficient curves and Mahalanobis distance diagnostics for selected Wolfenstein amplitudes at 100 MeV are shown in Fig.~\ref{fig10}.
In contrast to the features seen in Fig.~\ref{fig4} for $np$ scattering, it is $c_2$, instead of $c_0$, that is large for \Repp{A} and \Impp{A}. That can be easily understood since the $^3S_1-^3D_1$ channel, which receives a large contribution at leading order in $\chi$EFT due to the one-pion exchange, is absent. For the same reason, \Repp{A} does not show the strong increase at backward angles seen in \Renp{A}. Thus, it may not be surprising that the most significant coefficient is associated with the NLO term in the expansion, $c_2$. 

Indeed, for \Repp{A}, \Impp{A}, and \Repp{M} the N$^2$LO coefficient, $c_3$, is also as large or larger than $c_0$. These changes, as compared to the $np$ case, are likely because the $^1$P$_1$ and $^3$S$_1$-$^3$D$_1$ partial waves, which gave quite large contributions to the $np$ amplitudes at 100~MeV, are absent in the $pp$ case. 

In the $np$ case, we argued that for \Imnp{C}, the vector (spin-orbit) amplitude, our analysis {\it de facto} starts as NLO, since there is no contribution at LO. That is also the case here. And, as in the $np$ case, the NLO coefficient in \Impp{C} is markedly larger than all the other coefficients. 

In Fig.~\ref{fig11}, we see that, as with the truncation errors for $np$ scattering at 100~MeV, the 95\% interval obtained for the LO \Repp{A} only barely covers the higher-order and final results at forward angles. In the $pp$ case, the NLO errors also appear to be underestimated for \Repp{M}, especially at the smaller momentum transfers. The figures for the remaining tensor amplitudes are shown in Appendix~\ref{appendixB}.

Meanwhile, Fig.~\ref{fig12} shows that, at 25~MeV, 
the 95\% credibility interval at NLO comfortably encloses both the N$^4$LO and CD-Bonn results, as does the same interval at LO for \Repp{M} at 25~MeV. 

However, the LO uncertainty bands for both \Repp{A} and \Impp{A} significantly understate the size of the NLO correction at forward angles and only accommodate the full result once they become relatively large at backward scattering angles and momentum transfers $> 120$ MeV. 
This suggests that LO in the standard formulation of $\chi$EFT is not only insufficient when one considers higher-energy scattering or vector/tensor scattering amplitudes: it is not a particularly good starting point even in the case of $A_{pp}$ at low energies and momentum transfers.
It is also worth pointing out that the $\chi$EFT truncation uncertainty at N$^4$LO is very small: at forward angles, the red band in Fig.~\ref{fig12} is not much wider than the red line itself.
Low-energy scattering, therefore, represents quite a stringent test of a high-order $\chi$EFT interaction and the uncertainties assigned to it.  

\begin{figure}[h!]
\includegraphics[width=0.7\prcColumnWidth]{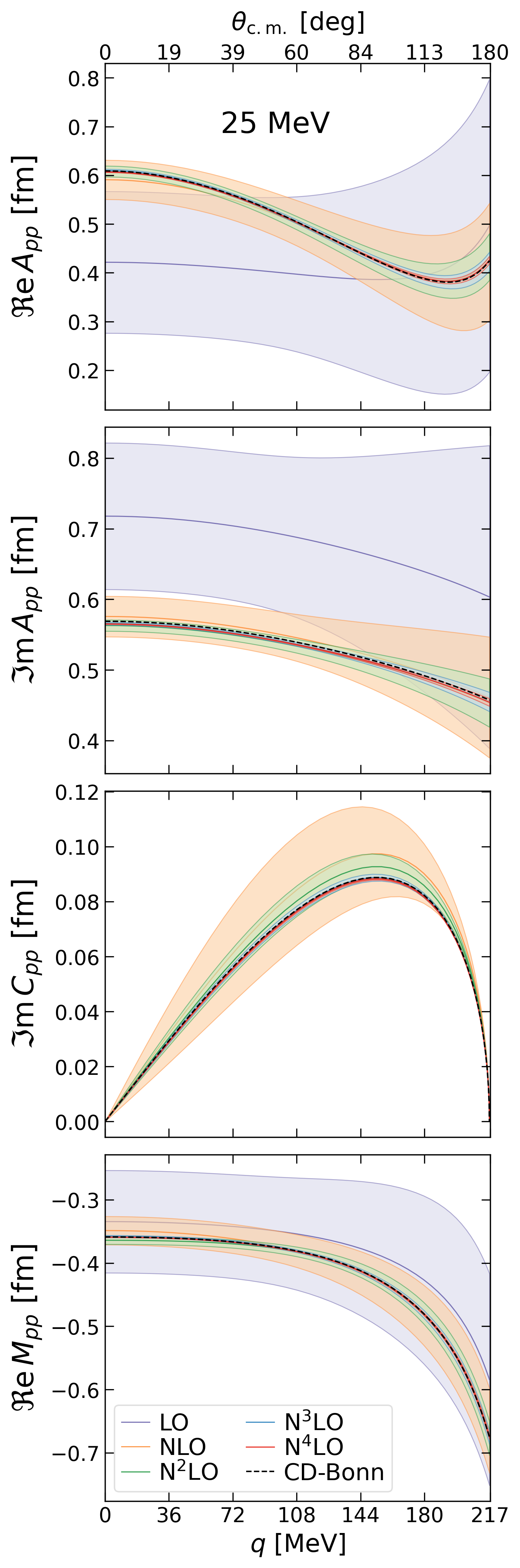}
\caption{The predictions and their corresponding 2$\sigma$ truncation error estimates obtained with different $\chi$EFT orders of the SCS interaction for $pp$ scattering Wolfenstein amplitudes at 25~MeV. These are generated via a process similar to that represented in Fig.~\ref{fig10}. The black dashed line results from a calculation based on the CD-Bonn potential.}
\label{fig12}
\end{figure}

\begin{figure}[h!]
\includegraphics[width=0.7\prcColumnWidth]{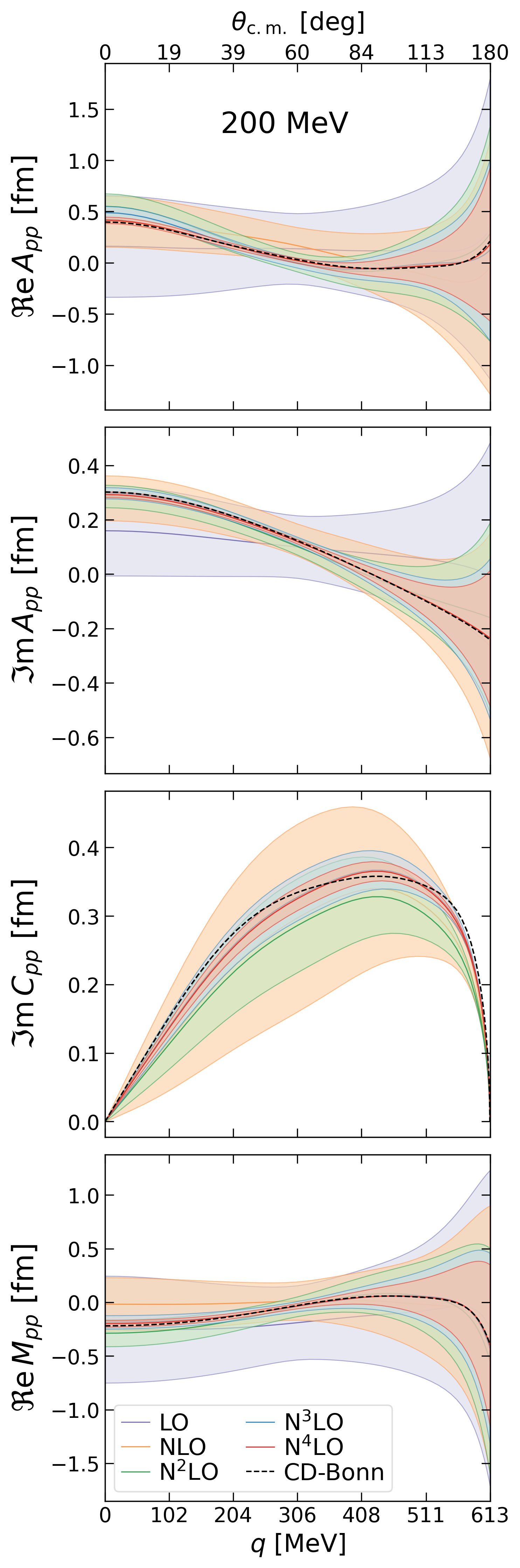}
\caption{The predictions and their corresponding 2$\sigma$ truncation error estimates obtained with different $\chi$EFT orders of the SCS interaction for $pp$ scattering Wolfenstein amplitudes at 200~MeV. These are generated via a process similar to the one represented in Fig.~\ref{fig10}. The black dashed line results from a calculation based on the CD-Bonn potential.}
\label{fig13}
\end{figure}

For the case of $E_{\rm lab}=200$ MeV, shown in Fig.~\ref{fig13}, the uncertainties obtained from the BUQEYE procedure for \Repp{A} and \Repp{M} have good coverage properties. Once again, the NLO shift in \Impp{A} at forward angles is right at the edge of the 95\% interval assigned to the LO prediction. The 95\% intervals derived for \Impp{C} at NLO and N$^2$LO do not envelop the higher-order errors or cover the CD-Bonn curve. As in the $np$ case, the shape of the CD-Bonn result for this observable seems somewhat different from that found even at the highest order computed with the SCS potential.

\begin{figure}[hbt]
\includegraphics[width=0.59\linewidth]{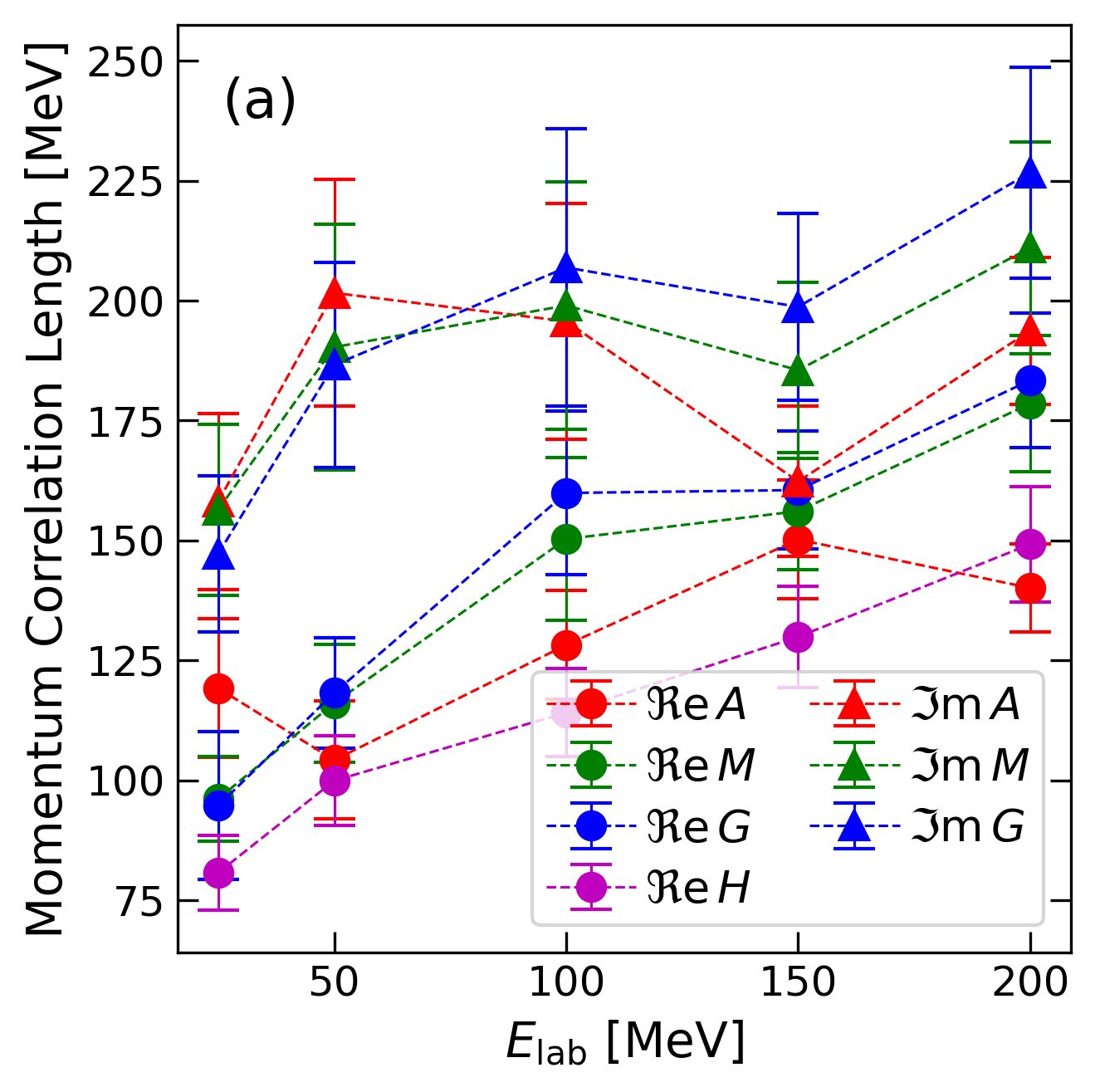}
\includegraphics[width=0.59\linewidth]{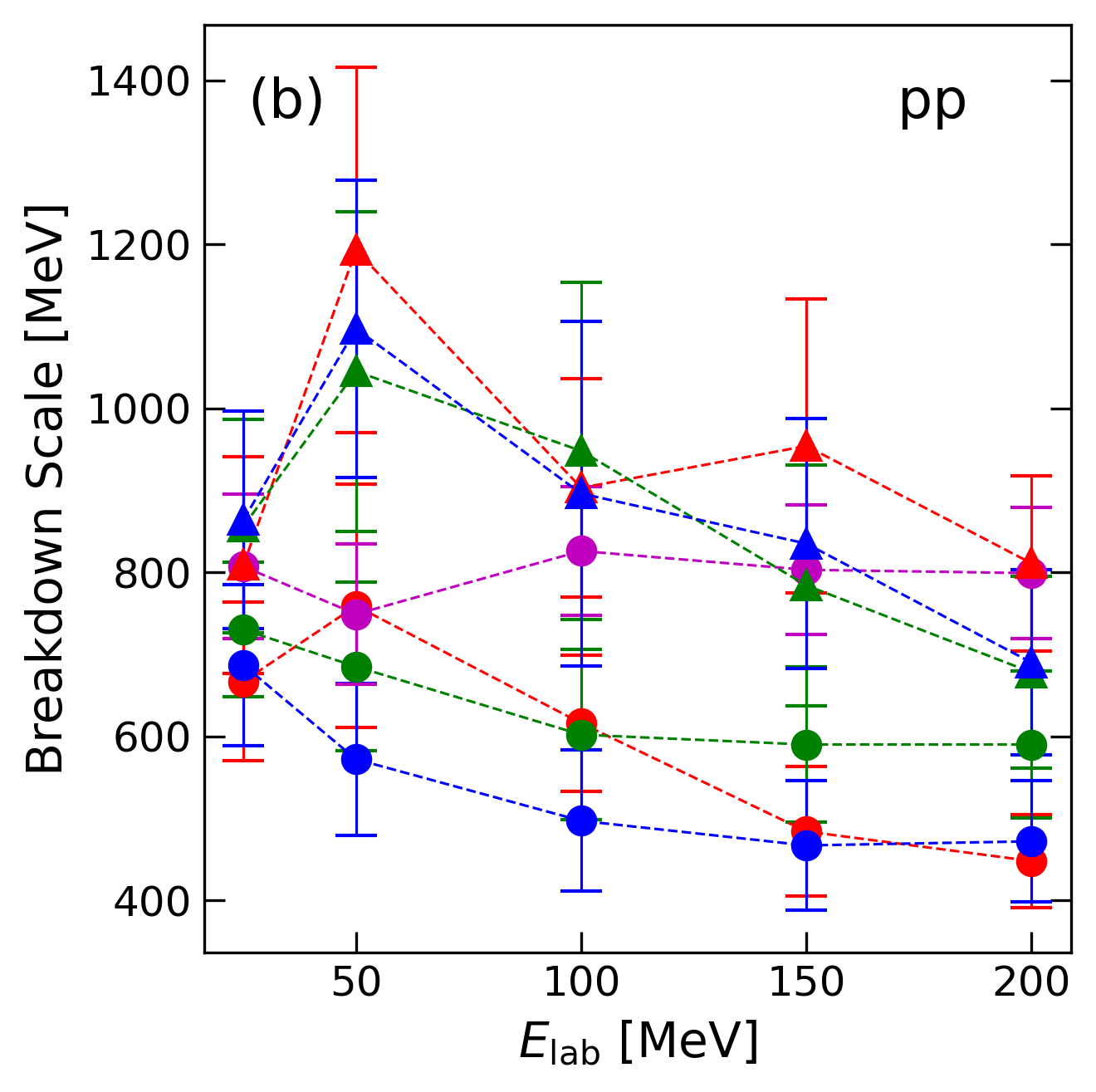}
\caption{Maximum {\it a posteriori} values for momentum correlation lengths, $\ell_q$, and $\chi$EFT breakdown scales, $\Lambda_b$, derived from selected $pp$ Wolfenstein amplitudes. The error bars denote the 1$\sigma$/68\% credibility interval.}
\label{fig14}
\end{figure}

Figure~\ref{fig14} is the analog of Fig.~\ref{fig9} for the $np$ case. It shows the 68\% intervals for both $\ell_q$ (top panel) and $\Lambda_b$ (bottom panel) at energies from 25 to 200 MeV for seven different Wolfenstein amplitudes. 

Here, we see some evidence in the real components of the tensor amplitudes for a rise in the momentum correlation length, $\ell_q$, as energy increases. We note, however, that for any given amplitude, the correlation length appears flat (within error bars) for $E_{\rm lab} \geq 100$ MeV. It also tends to be slightly larger than in the $np$ case. In the $pp$ case, the partial-wave structure of the Wolfenstein amplitudes is simpler, so there is more variability in the correlation lengths across the real and imaginary parts of $A$, $M$, $G$, and $H$. It is interesting to observe that $\Re{\rm e} \, H$ has the smallest $\ell_q$ for both $np$ and $pp$, while $\Im{\rm m} \, M$ exhibits the largest $\ell_q$ both here and in Fig.~\ref{fig9}.

Meanwhile, the breakdown scales, $\Lambda_b$, extracted at different energies are remarkably consistent for a given amplitude. But \Repp{A} and \Repp{G} exhibit consistently lower breakdown scales than we see in the imaginary parts of the amplitudes. The average MAP value for $\Lambda_b$ across all energies and all angles is 750 MeV. If, as we did for $np$, we exclude $\Re{\rm e} \, H$ from the average, this comes down to 740 MeV. This is reassuringly close to the average $\Lambda_b$ obtained from our $np$ analysis, but we reiterate the caveat that this averaging procedure is almost certainly not statistically consistent.


\section{Conclusions}
\label{sec:conclusions}

Order-by-order predictions of a well-behaved effective field theory should converge in a regular fashion towards a value given by the highest order calculated. In the present study, we followed the BUQEYE approach to EFT truncation errors by expressing different components of the NN amplitude in terms of dimensionless coefficients associated with each order of the EFT expansion. We used a decomposition of the NN amplitude into scalar, vector, and tensor operators: the Wolfenstein amplitudes for $np$ and $pp$ scattering. We examined the $\chi$EFT predictions for those obtained with the semi-local co-ordinate space (SCS) $\chi$EFT NN potential of Epelbaum, Krebs, and Mei\ss ner~\cite{Epelbaum:2014efa,Epelbaum:2014sza} with a cutoff of $R=0.9$~fm.

With appropriate choices for the EFT expansion parameter and the overall scale of the amplitudes, a Gaussian process can well describe the EFT coefficients of these amplitudes in the range of laboratory energies from 25 to 200 MeV. This energy range was chosen bearing in mind that S-waves mainly dominate at its lower end, while, for the highest energy we show, S-waves are no longer relevant.
Order-by-order convergence for the Wolfenstein amplitudes is thus overall consistent with the BUQEYE statistical model.
To be more specific, we needed to choose the reference value, $y_{\rm ref}$,
such that it captures the gross features of the amplitudes. For both the $np$ and $pp$ analyses, the imaginary part of the scalar $np$ amplitude, \Imnp{A}, at the corresponding laboratory kinetic energy is a good choice.
For the expansion parameter $Q$, we used the expression given in Eq.~(\ref{eq:Qchoice}), which is similar to the expression given in Millican {\it et al.}.

One output of this detailed analysis is a breakdown scale $\Lambda_b$ for $\chi$EFT. For the SCS interaction we analyzed, this turns out to be around 750~MeV.
It is remarkable that if one averages over the real and imaginary amplitudes, omits the vector amplitude $C$, and works only with the real tensor amplitudes, the average breakdown scale is roughly the same for the $np$ and $pp$ amplitudes. (The imaginary tensor amplitudes are, in general, much smaller than their real parts, which is one of the reasons we did not consider them in our analysis.)
We also extracted a momentum correlation length, which turns out to be quite similar for $np$ and $pp$ scattering and, intriguingly, relatively close to the physical pion mass. It has an average of about 140~MeV for $np$ scattering and a slightly higher value of 160~MeV for $pp$ scattering.

The order-by-order EFT predictions and corresponding $2\sigma$  truncation uncertainties at each order were computed, displayed, and examined for the chiral orders at $E_{\rm lab}=25$, $100$, and $200$ MeV. 
We find that the order-by-order truncation errors given by the BUQEYE analysis scheme work quite well when applied to the Wolfenstein amplitudes of NN scattering at NLO and beyond. Coefficients $c_2$ to $c_5$ have similar sizes and structures for the real parts of the amplitudes $A$, $M$, $G$, and $H$. 

The vector amplitude related to the spin-orbit force always needs special consideration since the $\chi$EFT expansion does not have a leading order contribution to this amplitude. 
Indeed, in general, we find that the leading order of $\chi$EFT, where the potential consists solely of one-pion exchange and S-wave counterterms, is a somewhat poor starting point for the regular order-by-order convergence pattern encoded in the BUQEYE model. After all, this leading-order result is relatively far from the highest-order and the ``experimental'' result---especially for $pp$ scattering. Particularly in the $pp$ case, most of the relevant contributions to NN scattering occur at NLO, not at LO, according to the standard organization of the $\chi$EFT potential. However, as already discussed, the effects in $\chi$EFT that occur beyond NLO for the SCS $R=0.9$ fm interaction produce a quite regular convergence pattern.

Analyzing order-by-order correlated truncation errors in terms of the invariant operator structure of the NN interaction thus gives a relatively concise picture of the convergence pattern of the $\chi$EFT. One-pion exchange plus momentum-indepedent counterterms is a natural starting point and can describe very low-energy observables that are dominated by S-waves. But, especially when entering the energy realm where P- and higher partial waves become important, it fails to describe NN scattering even within the large uncertainties assigned to it.


\begin{acknowledgments}
	This work was partly performed under the auspices of the U.S. Department of Energy under contract No.~DE-FG02-93ER40756. The authors acknowledge stimulating discussions with C. Drischler and R.~J. Furnstahl, and are grateful for C. Drischler's comments on the manuscript. B.~M. thanks A.~C. Semposki for technical advice while performing parts of the error analysis. The authors thank A. Nogga for sharing the codes that generate the $\chi$EFT interactions of the LENPIC collaboration.

\end{acknowledgments}

\clearpage


\begin{appendix}
\section*{Appendix}
\section{Additional tensor amplitudes for $\bm{np}$ scattering}
\label{appendixa}

\begin{figure}[h!]
\includegraphics[width=0.70\prcColumnWidth]{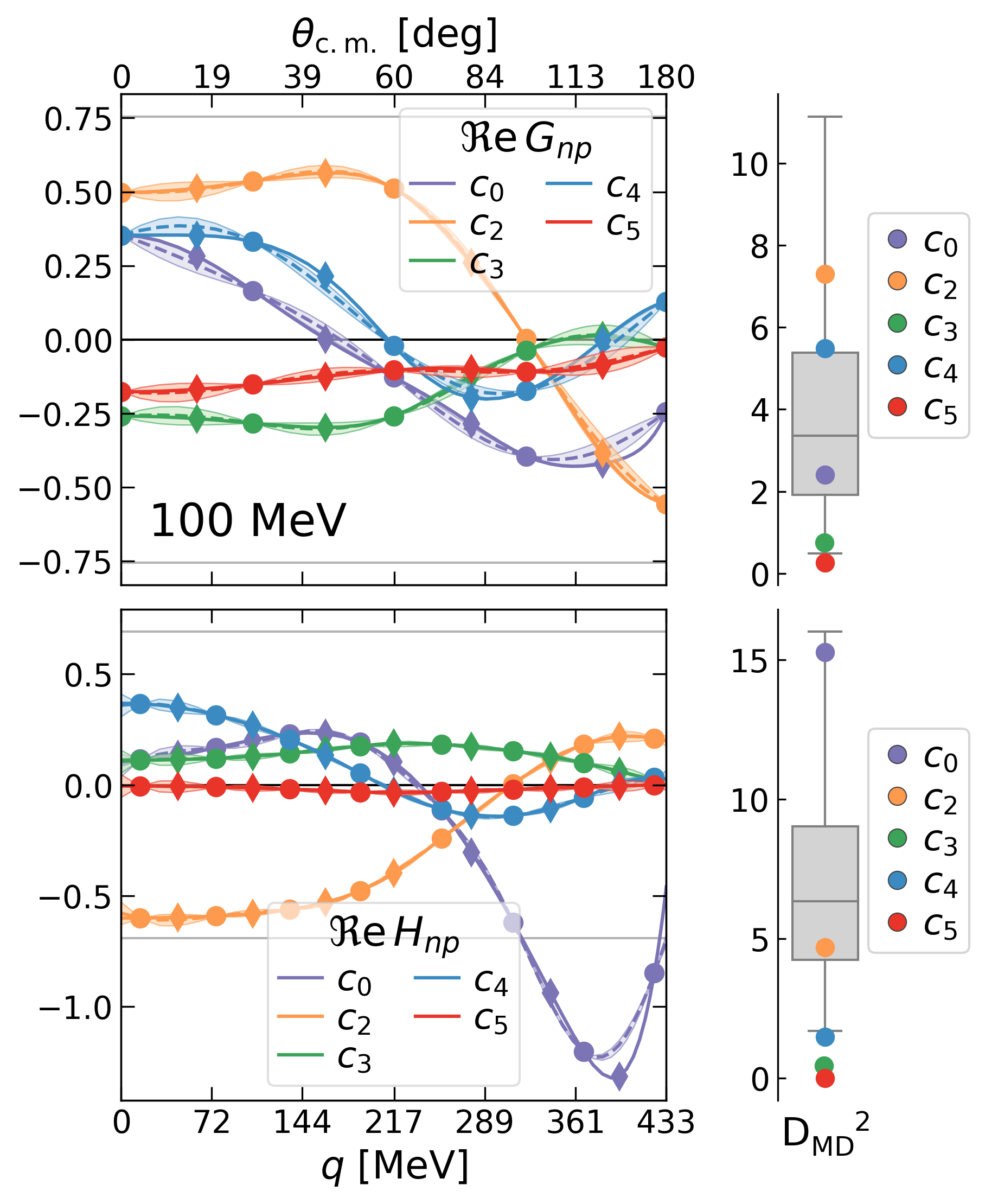}
\caption{Coefficient curves and Mahalanobis distance diagnostics for \Renp{G} and \Renp{H} at 100~MeV. For curves and legends, see Fig.~\ref{fig4}.
}
\label{fig15}
\end{figure}

\begin{figure}[h!]
\includegraphics[width=0.65\prcColumnWidth]{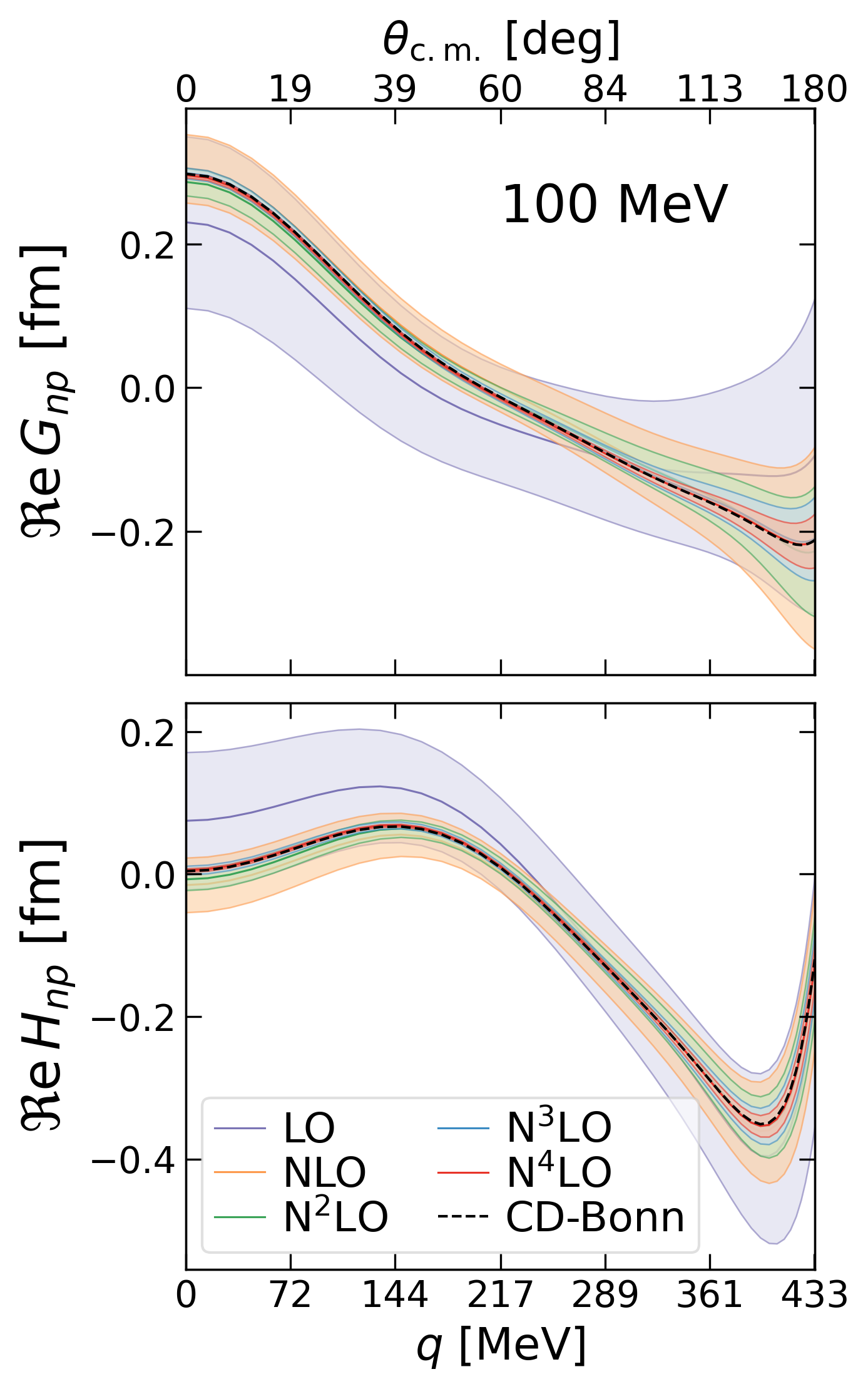}
\caption{The predictions and their corresponding 2$\sigma$ truncation error estimates for \Renp{G} and \Renp{H} at 100~MeV. For curves and legends, see Fig.~\ref{fig6}.
}
\label{fig16}
\end{figure}

\begin{figure}[b!]
\includegraphics[width=0.65\prcColumnWidth]{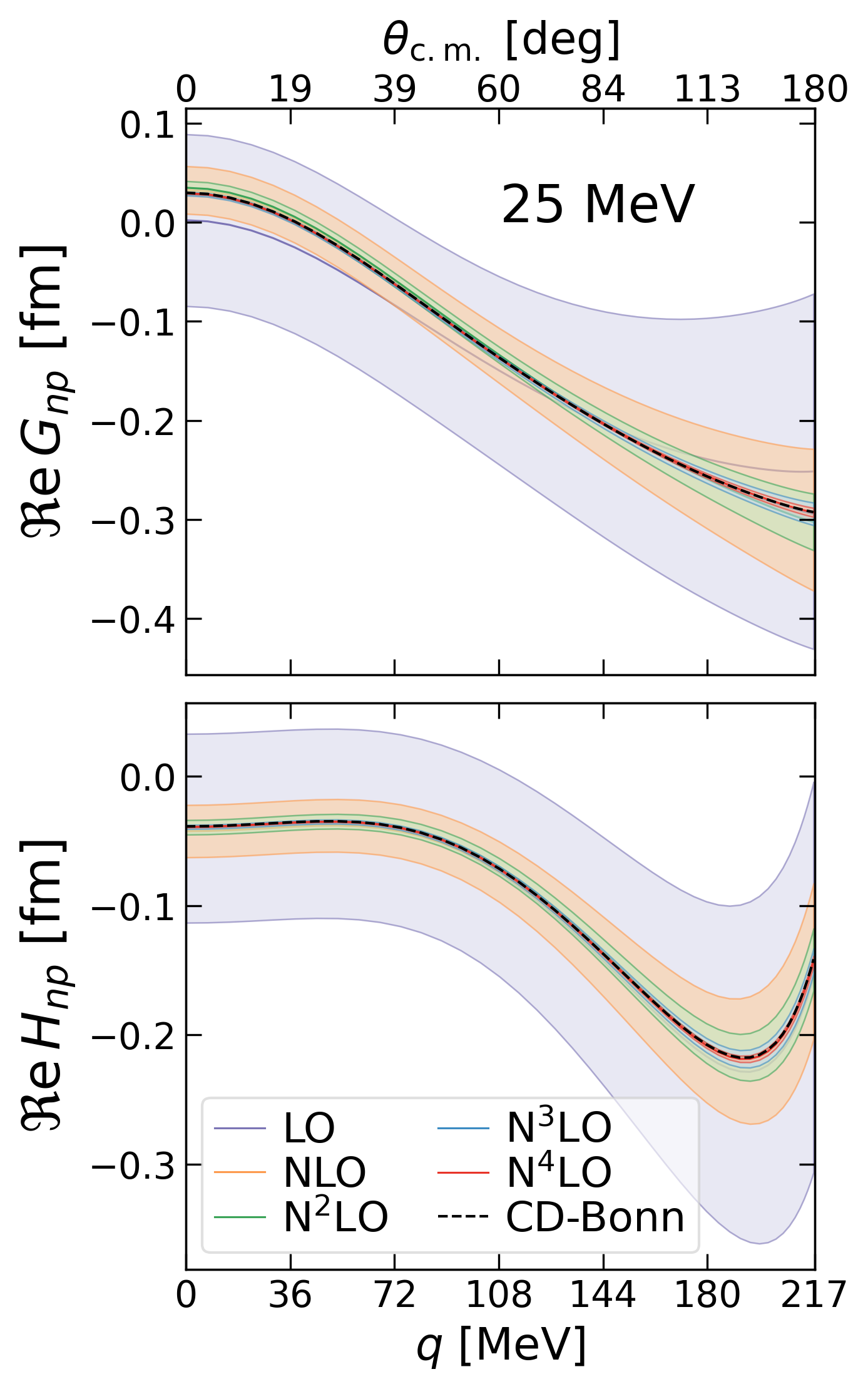}
\caption{The predictions and their corresponding 2$\sigma$ truncation error estimates for \Renp{G} and \Renp{H} at 25~MeV. For curves and legends, see Fig.~\ref{fig7}.
}
\label{fig17}
\end{figure}

\begin{figure}[b!]
\includegraphics[width=0.65\prcColumnWidth]{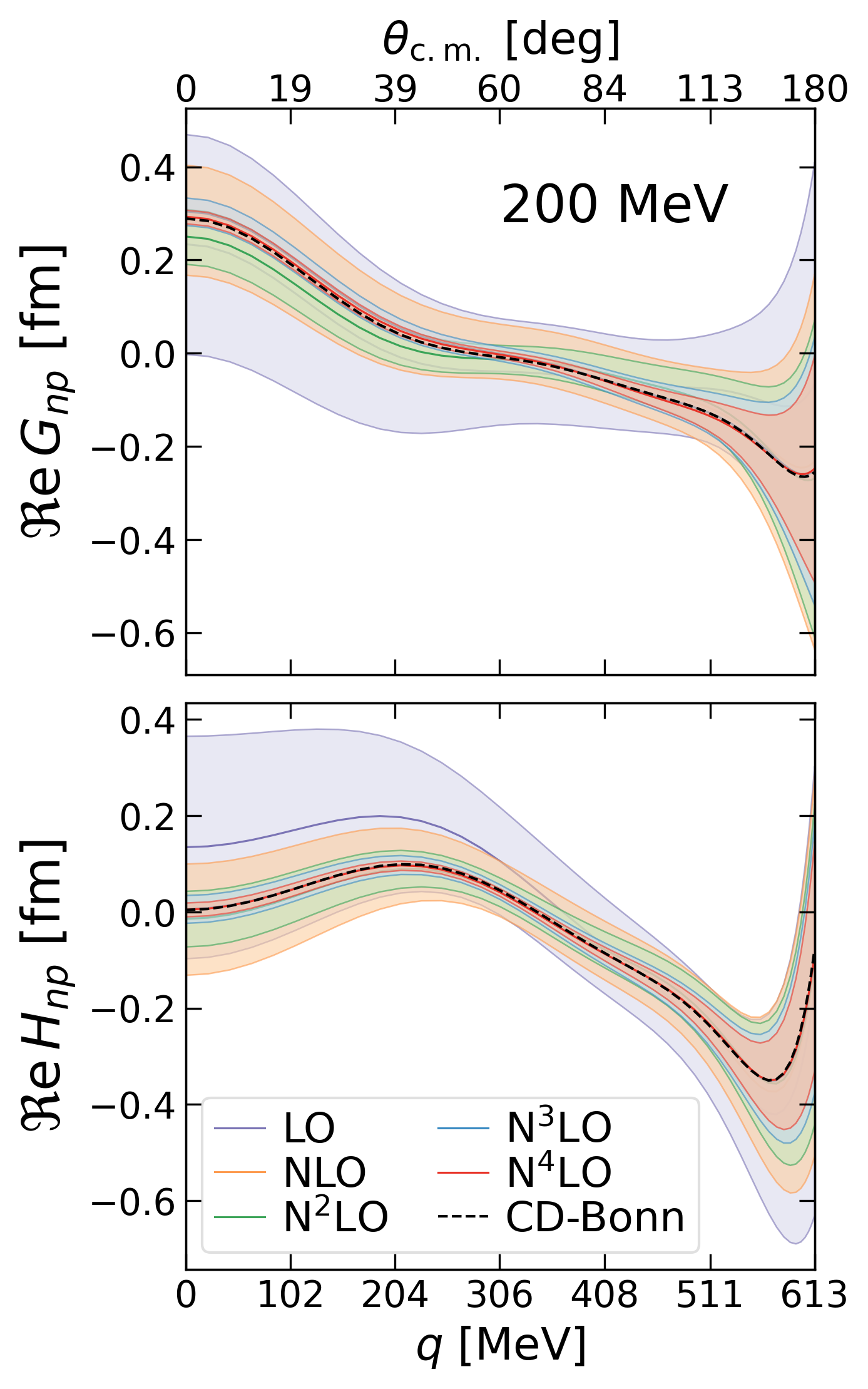}
\caption{The predictions and their corresponding 2$\sigma$ truncation error estimates for \Renp{G} and \Renp{H} at 200~MeV. For curves and legends, see Fig.~\ref{fig8}.
}
\label{fig18}
\end{figure}

\clearpage


\section{Additional tensor amplitudes for $\bm{pp}$ scattering}
\label{appendixB}

\begin{figure}[h!]
\includegraphics[width=0.70\prcColumnWidth]{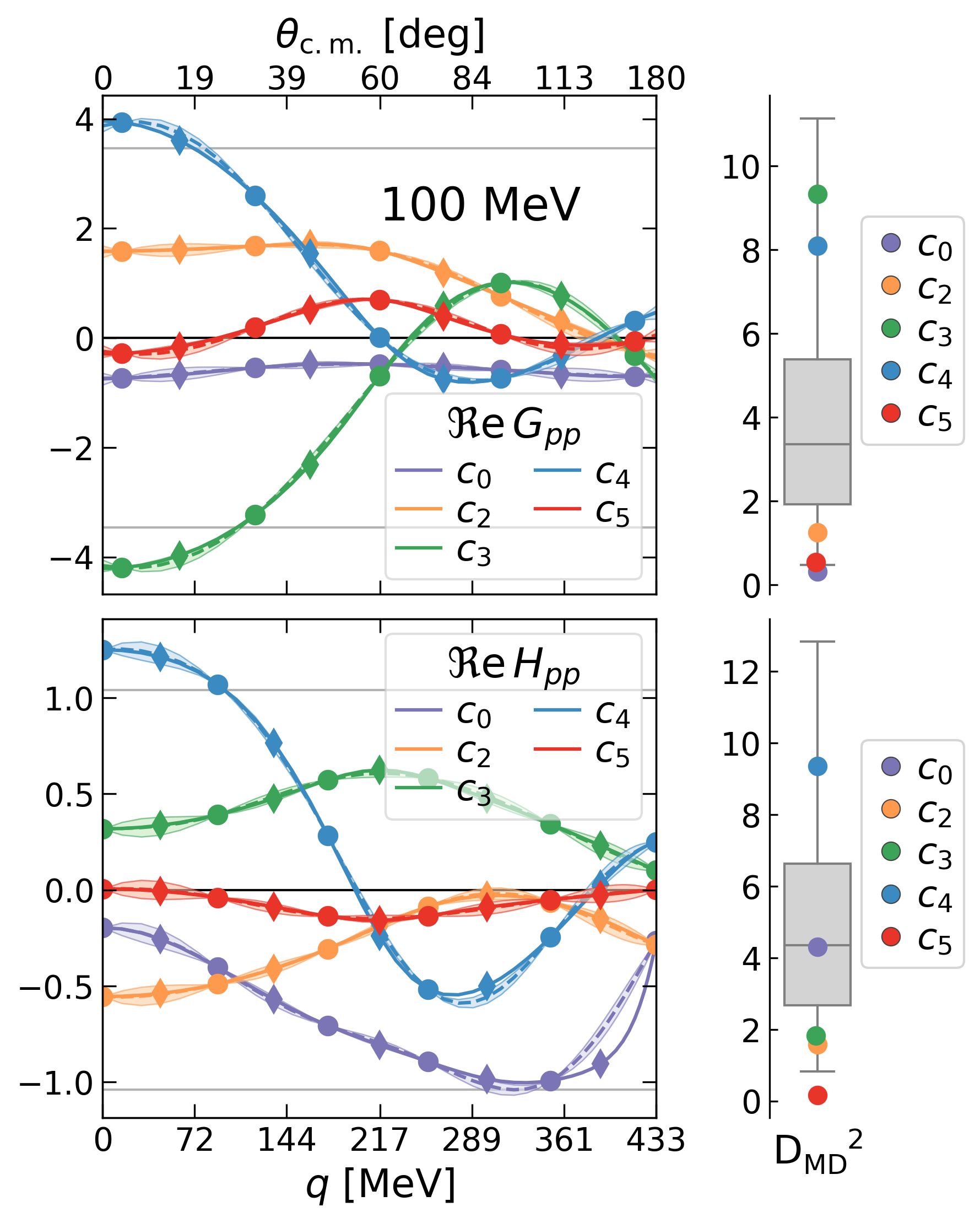}
\caption{Coefficient curves and Mahalanobis distance diagnostics for \Repp{G} and \Repp{H} at 100~MeV. For curves and legends, see Fig.~\ref{fig10}.
}
\label{fig19}
\end{figure}

\begin{figure}[h!]
\includegraphics[width=0.65\prcColumnWidth]{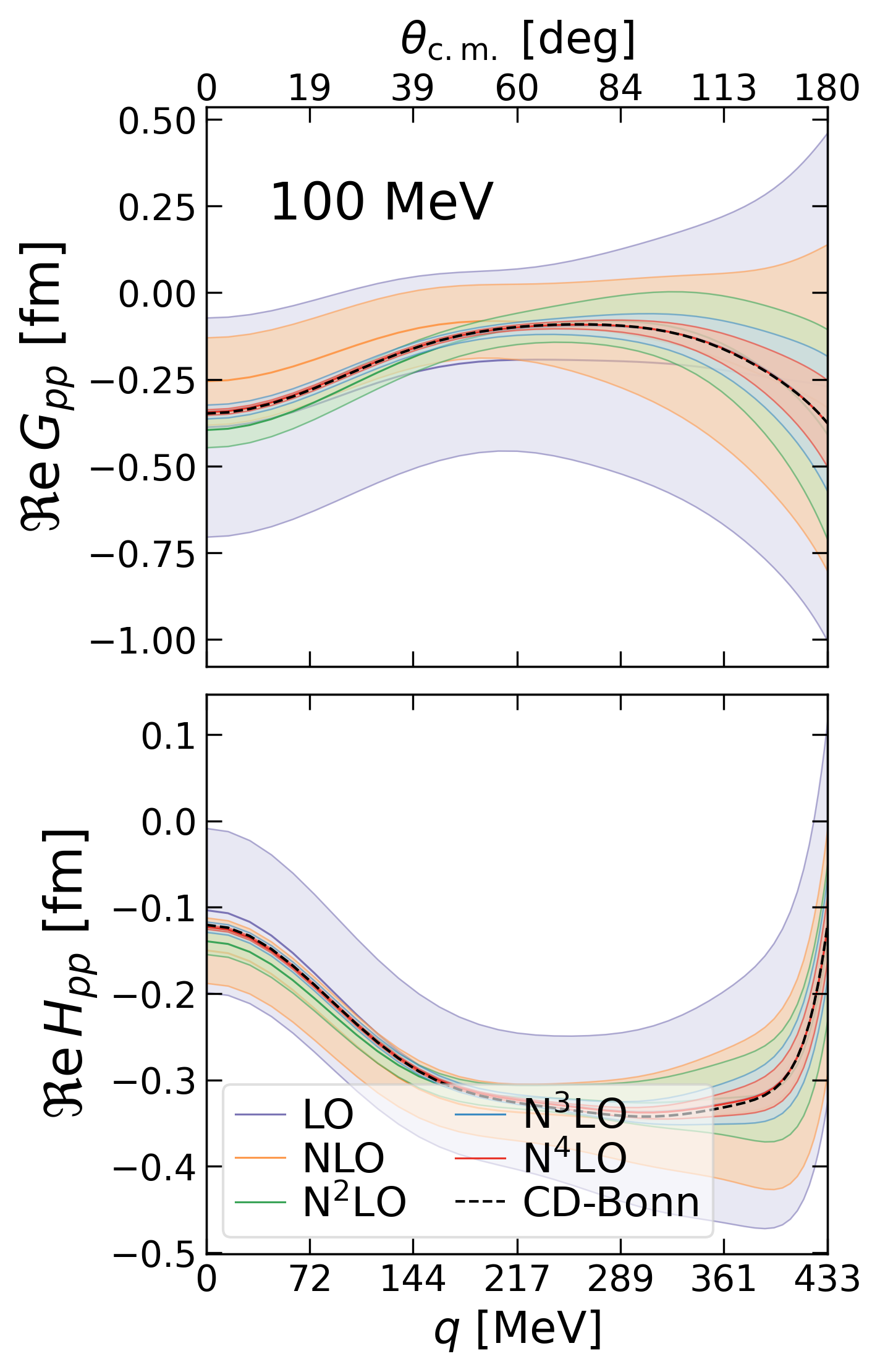}
\caption{The predictions and their corresponding 2$\sigma$ truncation error estimates for \Renp{G} and \Renp{H} at 100~MeV. For curves and legends, see Fig.~\ref{fig11}.
}
\label{fig20}
\end{figure}

\begin{figure}[b!]
\includegraphics[width=0.65\prcColumnWidth]{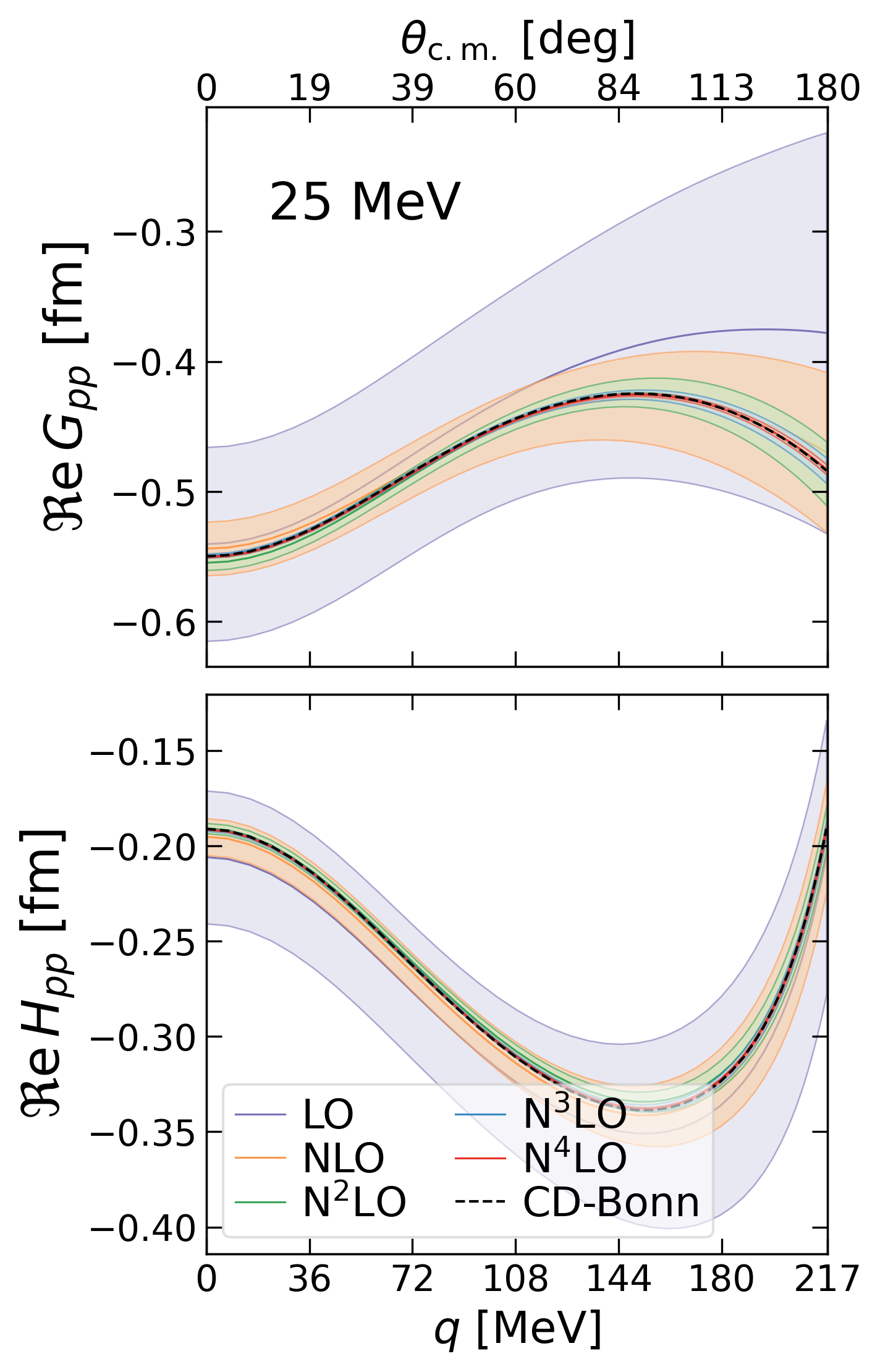}
\caption{The predictions and their corresponding 2$\sigma$ truncation error estimates for \Renp{G} and \Renp{H} at 25~MeV. For curves and legends, see Fig.~\ref{fig12}.
}
\label{fig21}
\end{figure}

\begin{figure}[b!]
\includegraphics[width=0.65\prcColumnWidth]{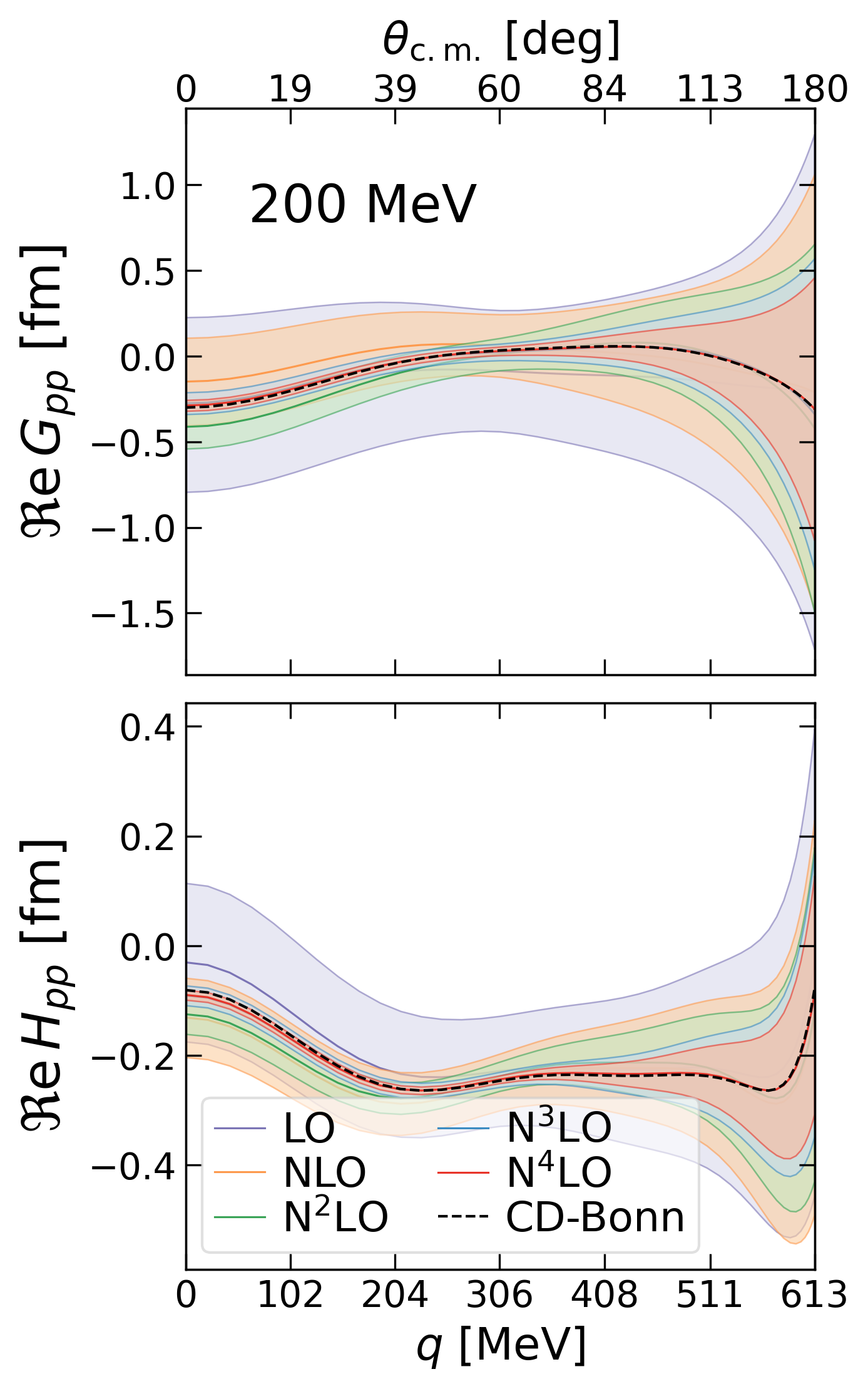}
\caption{The predictions and their corresponding 2$\sigma$ truncation error estimates for \Renp{G} and \Renp{H} at 200~MeV. For curves and legends, see Fig.~\ref{fig13}.
}
\label{fig22}
\end{figure}

\end{appendix}
\clearpage

\bibliography{reactions,clusterpot,ncsm,bayesian_refs}

\end{document}